\documentclass[twoside]{ahp}

\usepackage{kempe_h-corr}
\usepackage{amssymb,amsmath}
\usepackage{epsfig}

\newcommand{\eps}{\varepsilon}
\newcommand{\ket}[1]{|#1\rangle}

\newcommand{\be}{\begin{equation}}
\newcommand{\ee}{\end{equation}}
\newcommand{\la}{\langle}
\newcommand{\ra}{\rangle}
\newcommand{\mtwo}[4] {\left( \begin{array}{cc}{#1}&{#2}\\{#3}&{#4}\end{array} \right)}

\begin{document}

\title{Approaches to Quantum Error Correction}
\author{Julia {\sc Kempe}\\CNRS \& LRI\\Laboratoire de Recherche Informatique\\B\^at. 490,
Universit\'e de Paris-Sud\\91405 Orsay Cedex\\ France}

 \maketitle

We have persuasive evidence that a quantum computer would have extraordinary power. But will we ever be able
to build and operate them? A quantum computer will inevitably interact with its environment, resulting in
decoherence and the decay of the quantum information stored in the device. It is the great technological (and
theoretical) challenge to combat decoherence. And even if we can suitably isolate our quantum computer from
its surroundings, errors in the quantum gates themselves will pose grave difficulties. Quantum gates (as
opposed to classical gates) are unitary transformations chosen from a {\em continuous} set; they cannot be
implemented with perfect accuracy and the effects of small imperfections in the gates will accumulate,
leading to an eventual failure of the computation. Any reasonable correction-scheme must thus protect against
small unitary errors in the quantum gates as well as against decoherence. Furthermore we must not ignore that
the correction and recovery procedure itself can introduce new errors; successful fault-tolerant quantum
computation must also deal with this issue.

The purpose of this account is to give an overview of the main approaches to quantum error correction. There exist
several excellent reviews of the subject, which the interested reader may consult (see
\cite{Preskill:98a},\cite{Preskill:99a}, \cite{Nielsen:book}, \cite{Kitaev:book}, \cite{Steane:99a,Steane:01a} and more
recently \cite{Gottesman:05a}).

\section{Introduction}

\begin{center}
{\em ``We have learned that it is possible to fight entanglement with entanglement."\\John Preskill, 1996}
\end{center}

In a ground breaking discovery in 1994, Shor \cite{Shor:94a} has shown that quantum computers, if built, can factor
numbers efficiently. Since then quantum computing has become a burgeoning field of research, attracting theoreticians
and experimentalists alike, and regrouping researchers from fields like computer science, physics, mathematics and
engineering. One more reason for the enormous impetus of this field is the fact that by the middle of 1996 it has been
shown how to realize {\em fault-tolerant} quantum computation. This was not at all obvious; in fact it was not even
clear how any form of quantum error-correction could work.
 Since then many new results
about the power of quantum computing have been found, and the theory of quantum fault-tolerance has been developed and
is still developing now.

In what follows we will give a simple description of the elements of quantum error-correction and quantum
fault-tolerance. Our goal is to convey the necessary intuitions both for the problems and their solutions. In no way
will we attempt to give the full and formal picture. This account is necessarily restricted with subjectively chosen
examples and approaches and does not attempt to describe the whole field of quantum fault-tolerance, which has become a
large subfield of quantum information theory of its own.

The structure of this account is the following. First we will describe why quantum error correction is a non-trivial
achievement, given the nature of quantum information and quantum errors. Then we will briefly review the main features
of a quantum computer, since, after all, this is the object we want to protect from errors, and it is also the object
which will allow us to implement error-correction. Then we will give the first example of a quantum error-correcting
code (the Shor-code), followed by other error correction mechanisms. We proceed to outline the elements of a full
fledged fault-tolerant computation, which works error-free even though all of its components can be faulty. We mention
alternative approaches to error-correction, so called error-avoiding or decoherence-free schemes. We finish with an
outlook on the future. We will try to keep technical details and generalizations to a minimum; the interested reader
will find more details, as well as suggestions for further reading, in the appendix.

\section{The subtleties of quantum errors}

\begin{center}
{\em ``Small errors will accumulate and cause the computation to go off track."}\\ {\em Rolf Landauer, 1995, in ``Is
quantum mechanics useful?"}
\end{center}

A quantum machine is far more susceptible to making errors then classical digital machines.

Not only is a quantum system more prone to decoherence resulting from unwanted interaction between the quantum system
and its environment, but also, when manipulating quantum information we can only implement the desired transformation
up to a certain precision. Until 1995 it was not clear at all if and how quantum error correction could work.

The second big breakthrough towards quantum computing (after Shor's algorithm) was the insight that quantum noise can
be combatted or that quantum error protection and correction is possible. The first big step in this direction was
again made by Peter Shor in his {\em ``Scheme for reducing decoherence in a quantum memory''} in 1995 \cite{Shor:95a}.

This was indeed an amazing piece of work: the difficulties facing the introduction of classical error-correction ideas
into the quantum realm seemed formidable. In fact there was a large number of reasons for pessimism. Let us cite but a
few of the apparent obstacles:

(1) There is a hugely successful theory of classical error correction which allows to protect against classical errors.
However, classical errors are {\em discrete} by their nature. In the most common case where the information is encoded
into a string of bits, the possible errors are bit-flips or erasures. A quantum state is a priori continuous, and hence
also the error is continuous. Similarly, quantum {\em operations} are continuous by their nature, and will necessarily
only be implemented with a certain precision, but never exactly. As noted by Landauer \cite{Landauer:95a}, small errors
can accumulate over time and add up to large, uncorrectable errors. Moreover it is not clear how to adapt the discrete
theory of error correcting codes to the quantum case.

(2) A second objection is that to protect against errors, the information must be encoded in a redundant way. However,
the quantum no-cloning theorem \cite{Dieks:82a,Wootters:82a}, which follows directly from the linearity of quantum
mechanics, shows that it is impossible to copy an unknown quantum state. How then can the information be stored in a
redundant way?

(3) Another point is the following: in order to correct an error, we need to first acquire some information about the
nature and type of error. In other words we need to observe the quantum system, to perform a measurement. But any
measurement collapses the quantum system and might destroy the information we have encoded in the quantum state. How
then shall we extract information about the error without destroying the precious quantum superposition that contains
the information?

Many researchers in the field were pessimistic about the prospects of
error-correction and Shor's result came as a great surprise to many. All the
initial objections let us appreciate the elegance of the solution even more.
But before giving the key ingredients, we need to briefly review the object we
want to protect from errors, the actual quantum machine.

\section{What is a quantum computer?}

\begin{center}
{\em ``The more success the quantum theory has, the sillier it looks." \\Albert Einstein, 1912}
\end{center}

There are nearly as many proposals for the hardware of a quantum computer today, as there are experimental quantum
physicists. The ultimate shape and function of a quantum computer will depend on the physical system used, be it
optical lattices, large molecules, crystals or silicon based architectures. Nonetheless, each of these implementations
have some key elements in common, since they all implement the quantum computing model.

What are the key ingredients of a quantum computer? A quantum computer is a machine that processes basic computational
units, so called {\em qu}bits, quantum two-level systems. (Although there might be quantum machines which process
higher-dimensional quantum systems, we will restrict ourselves for simplicity to the case of two-level systems.) Qubits
are two-dimensional quantum states spanned by two basis-states, which we conventionally call $\ket{0}$ and $\ket{1}$,
alluding to the classical bits of a standard computer. Hence the general state of a qubit is
 \[ \ket{\psi}=\alpha \ket{0} + \beta \ket{1}  \quad \quad |\alpha|^2 + |\beta|^2 = 1,\]
 where $\alpha$ and $\beta$ are complex numbers.
In each implementation of a quantum computer these basis states $\ket{0}$ and $\ket{1}$ need to be identified; they
usually correspond to two chosen states of a larger system. For any quantum computation, {\em fresh} qubits have to be
supplied in a known state, which is usually taken to be the $\ket{0}$ state.

\begin{figure}[h]
 \center{\epsfxsize=1.5in \epsfbox{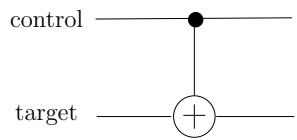}}
  \caption{\small The top qubit is the {\em control} qubit. If it is
in the state $\ket{0}$, then the {\em target} qubit stays unchanged, if it is in the state $\ket{1}$, the target qubit
is flipped from $\ket{0}$ to $\ket{1}$ and vice versa.}
 \label{pic1}
\end{figure}

A quantum computer implements a {\em unitary} transformation on the space of several qubits, as consistent with the
laws of quantum mechanics. However, in the context of computation, each unitary is decomposed into {\em elementary}
gates, where each gate acts on a small number of qubits only. These elementary gates constitute a universal gate set,
which allows to implement any unitary operation on the set of qubits. There are several universal gate sets, but we
will mention only two, which are relevant for what follows. The first such set is continuous, and consists of all
one-qubit unitaries, together with the controlled NOT or CNOT. The action of the CNOT on the basis states
$\ket{00},\ket{01},\ket{10},\ket{11}$ is as follows
 \[ CNOT = \left( \begin{array}{cccc} 1 & 0 & 0 & 0 \\ 0 & 1 & 0 & 0 \\ 0 & 0 & 0 & 1 \\ 0 & 0 & 1 & 0 \end{array}
 \right).\]
In quantum circuit design it is often depicted as in Fig. \ref{pic1}.

It is possible to implement any unitary operation by a sequence of CNOT and single qubit unitary operations on the
qubits.

The second set of universal gates is {\em discrete}. It contains the gates $H$,
$\pi/8$, $Z$ and $CNOT$. The first three gates are single qubit gates. $H$ is
called the {\em Hadamard} transform, $\pi/8$ is a phase gate and $Z$ is known
to physicists as one of the Pauli matrices $\sigma_z$\footnote{Note that the
$Z$ gate is not necessary for universality, as it can be generated from the
other gates in the set. However, it is often included for convenience, as it
becomes necessary in many fault-tolerant gate sets.}. On the basis $\ket{0}$,
$\ket{1}$, they act as
 \[ H =  \frac{1}{\sqrt{2}}\left( \begin{array}{cc} 1 & 1 \\ 1 & -1 \end{array} \right) \quad \quad \pi/8 =\left(
 \begin{array}{cc} e^{i \frac{\pi}{8}} & 0 \\ 0 & e^{-i \frac{\pi}{8}} \end{array} \right)
  \quad \quad Z = \left( \begin{array}{cc} 1 & 0
 \\ 0 & -1 \end{array} \right). \]
All experimental proposals, in one way or another, demonstrate the ability to induce the transformations corresponding
to this (or some other universal) gate set. Note that it is absolutely crucial to implement the two-qubit gate CNOT (or
some other suitable two-qubit gate), as single qubit operations alone are clearly not universal.

This gate set is discrete, it contains only four gates. However, this comes at a price. It is not in general possible
to implement any unitary transformation with a sequence drawn from this gate set. But it is possible to {\em
approximate} any unitary to arbitrary accuracy using gates from this set. (Here accuracy is measured as the spectral
norm of the difference between the desired unitary matrix and the actually implemented unitary matrix.) This is good
enough for our purposes.

The last ingredient of a quantum computer is the {\em read-out}, or measurement. At the end of the day, when we want to
extract the result of the quantum computation, we need to observe the quantum system, to gain information about the
result.

In general one assumes that each qubit (or the qubit carrying the result of the computation) is measured in some basis.
The classical result represents the outcome of the computation.

\begin{figure}[h]
 \center{\epsfxsize=2.5in \epsfbox{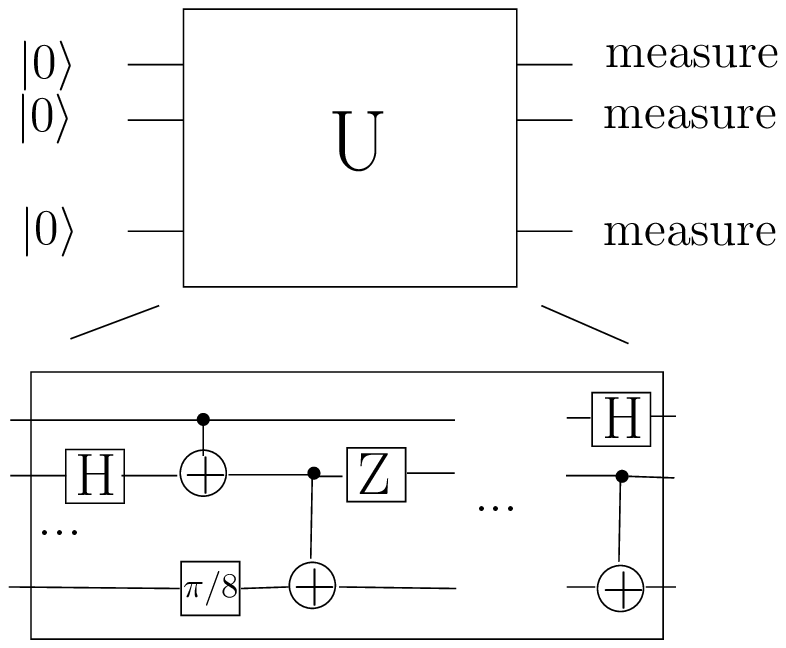}}
  \caption{\small A quantum computer, schematically. Fresh qubits, initialized in the state $\ket{0}$, are supplied as
  the input to the unitary transform $U$. $U$ is composed of elementary gates affecting at most $2$ qubits. At the end
  of the computation the qubits are measured.}
 \label{pic2}
\end{figure}

Schematically, then, a quantum computer looks like in Fig. \ref{pic2}.

\section{What is a quantum error?}

 \begin{center}
 {\em ``Had I known that we were not going to get rid of this damned quantum jumping,
 I never would have involved myself in this business!''\\Erwin Schr\"odinger}
 \end{center}

Quantum computers are notoriously susceptible to quantum errors, and this is
certainly the reason we did not yet succeed in building a scalable model. The
problem is that our quantum system is inevitably in contact with a larger
system, its {\em environment}. Even if we make heroic efforts to isolate a
quantum system from its environment, we still have to manipulate the
information inside it in order to compute, which again will introduce errors.
Where it not for the development of methods of quantum error correction, the
prospects for quantum computing technology would not be bright. In order to
describe quantum error correction we need to get a clear picture of what the
noise processes affecting our machine are.

But how do we describe a quantum error?

Let us study the example of a single qubit. This qubit might undergo some random unitary transformation, or it might
decohere by becoming entangled with the environment. In general it will undergo some unitary transformation in the {\em
combined} system of qubit and environment. Let us call $\ket{E}$ the state of the environment before the interaction
with the qubit. Then the most general unitary transformation on system and environment can be described as
 \begin{align*}
 U: \quad & \ket{0} \otimes \ket{E} \longrightarrow \ket{0} \otimes \ket{E_{00}} + \ket{1} \otimes \ket{E_{01}}\nonumber \\
    & \ket{1} \otimes \ket{E} \longrightarrow \ket{0} \otimes \ket{E_{10}} + \ket{1} \otimes \ket{E_{11}}.
 \end{align*}
Here $\ket{E_{ij}}$ represent not necessarily orthogonal or normalized states of the environment, with the only
constraint that the total evolution be unitary. The unitary $U$ {\em entangles} our qubit with the environment.
Potentially, this entanglement will lead to decoherence of the information stored in the qubit.

Suppose now the qubit is in the state $\alpha \ket{0} + \beta \ket{1}$\footnote{Of course our qubit could be part of a
larger quantum state of several qubits. It might be entangled with other qubits which are unaffected by errors. So the
coefficients $\alpha$ and $\beta$ need not be numbers, they can be states that are orthogonal to both $\ket{0}$ and
$\ket{1}$.}. Now if the qubit is afflicted by an error, it evolves as
 \begin{align}\label{eq:error}
\left( \alpha\ket{0} + \beta \ket{1} \right) \otimes \ket{E} \quad  \longrightarrow \quad & \alpha \left(\ket{0}
\otimes \ket{E_{00}} + \ket{1} \otimes \ket{E_{01}}\right) + \beta \left( \ket{0} \otimes \ket{E_{10}} + \ket{1}
\otimes \ket{E_{11}} \right)  \nonumber \\
=& \left( \alpha \ket{0} + \beta \ket{1} \right) \otimes \frac{1}{2} \left( \ket{E_{00}}+\ket{E_{11}} \right)
\quad \quad identity\nonumber \\
+& \left( \alpha \ket{0} - \beta \ket{1} \right) \otimes \frac{1}{2} \left( \ket{E_{00}}-\ket{E_{11}} \right)
\quad \quad phase \, flip \nonumber \\
+& \left( \alpha \ket{1} + \beta \ket{0} \right) \otimes \frac{1}{2} \left( \ket{E_{01}}+\ket{E_{10}} \right)
\quad \quad bit \, flip\nonumber \\
+& \left( \alpha \ket{1} - \beta \ket{0} \right) \otimes \frac{1}{2} \left( \ket{E_{01}}-\ket{E_{10}} \right) \quad
\quad bit \, \& \, phase \, flip.
 \end{align}
Intuitively, we may interpret this expansion by saying that one of four things
happens to the qubit: nothing, a bit flip, a phase flip or a combination of bit
flip and phase flip. This will be made more precise in the next section, where
we see that quantum error correction will include a {\em measurement} of the
error, collapsing the state into one of the four possibilities above. This way,
even though the quantum error is continuous, it will become {\em discrete} in
the process of quantum error correction. We will denote the four errors acting
on a qubit as
 \be \label{eq:Pauli}
\underbrace{I=\mtwo{1}{0}{0}{1}}_{identity} \quad \quad \underbrace{X=\mtwo{1}{0}{0}{-1}}_{phase \, flip} \quad \quad
\underbrace{Z=\mtwo{0}{1}{1}{0}}_{bit \, flip} \quad \quad \underbrace{Y=XZ=\mtwo{0}{-1}{1}{0}.}_{bit \, \& \, phase \,
flip}
 \ee
These four matrices form the so called {\em Pauli group}. Another way of saying the above  is to realise that these
four errors span the space of unitary matrices on one qubit, i.e. any matrix can be expressed as a linear combination
of these four matrices (with complex coefficients). If we trace out the environment (average over its degrees of
freedom, see App. \ref{sec:OSR}), the resulting operator can be expanded in terms of the Pauli group, we can attach a
probability to each Pauli group element. Often the analysis of fault-tolerant architectures is simplified by assuming
that the error is a random non-identity Pauli matrix with equal probability $\eps/3$, where $\eps$ is the {\em error
rate}.

We now make a crucial assumption: that the error processes affecting different
qubits are {\em independent} from each other. A quantum error correcting code,
then, will be such that it can protect against these four possible errors. Once
the error has become discrete it is much more obvious how to apply and extend
classical error correction codes, which are able to protect information against
a bit flip.

We have so far only analyzed errors due to decoherence, but have neglected errors due to imperfections in the gates, in
the measurement process and in preparation of the initial states. All these operations can be faulty. A natural
assumption is again that these imperfections are independent of each other. In a similar fashion as before we can
discretize the errors in a quantum gate. We can model a faulty gate by assuming that is is a perfect gate, followed by
an error. For a one-qubit gate this error is the same as given in Eq. (\ref{eq:error}). For a two-qubit gate we assume
that both qubits undergo possibly correlated decoherence. Similar reasoning as in Eq. (\ref{eq:error}) shows, that in
that case the error is a linear combination of $16$ possible errors, resulting from all combinations of the errors in
Eq. (\ref{eq:Pauli}) on both qubits. Again, often the additional assumption is made that all $15$ non-identity errors
appear with equal probability $\eps_2/15$, where $\eps_2$ is the two-qubit gate error rate. In a similar fashion we
will deal with measurement and state preparation errors.

Note that our analysis of the error is somewhat simplified. Several tools have been developed to study quantum
decoherence and quantum noise. Some of these formalisms are described in more detail in App. \ref{sec:decoherence}. As
already mentioned, in order to give methods for quantum error correction, some assumptions about the nature of the
noise have to be made. In one of the common models of noise in a quantum register it is assumed that each qubit
interacts {\em independently} with the environment in a Markovian fashion\footnote{This means that the environment
maintains no memory of the errors, which are thus {\em uncorrelated} in {\em time} and qubit {\em location}.}; the
resulting errors are single qubit errors affecting each qubit independently at random. More details on models of
quantum noise are given in App. \ref{sec:error-model}.

\section{The first error correction mechanisms}
\begin{center}
{\em ``Correct a flip and phase - that will suffice. \\If in our code another error's bred,\\ We simply measure it,
then God plays dice,\\ Collapsing it to X or Y or Zed."\\ Daniel Gottesman, in "Error Correction Sonnet"}
\end{center}

We have seen how entanglement with the environment can cause errors that result in a complete loss of the quantum
information. However, entanglement will also allow us to {\em protect} the information in a non-local way. If we
distribute the information over several qubits in a way that it cannot be accessed by measuring just a few of the
qubits, then by the same token it cannot be damaged if the environment interacts with just a few of the qubits.

A marvellous machinery has been developed in the classical world to protect classical information, the theory of error
correcting codes. The simplest possible such code is the {\em repetition} code: each bit is replaced by three of its
copies:
 \be
{\cal C}: \quad 0 \longrightarrow 000 \quad \quad \quad 1 \longrightarrow 111.\nonumber
 \ee
This code clearly protects against one bit flip error. If a bit is flipped, we can still decode the information by
majority voting. Only if two bit flips happen we will be unable to correctly decode the information. But if we assume
that the probability of a bit flip is $\eps$ and independent on each bit, then the probability that we cannot correct a
bit flip is $3\eps^2(1-\eps)+\eps^3$ (there are three possible ways to have two bit flips and one way to have three bit
flips). If we would not encode the information at all the error probability is $\eps$, so as long as $\eps < 1/2$ we
gain by encoding.

But how can we extend this idea to the quantum setting? There is no way to copy quantum information. There are not only
bit flip, but also phase flip errors (and combinations of both). And moreover a measurement for majority vote will
cause disturbance.

\begin{figure}[h]
 \center{\epsfxsize=2.5in \epsfbox{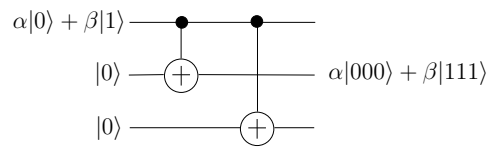}}
  \caption{\small The CNOTs flip the target qubit if the first qubit is in the state $\ket{1}$. Note that the
  transformation does not {\em copy} the state of the first qubit to the other two qubits, but rather implements the
  transformation of Eq.(\ref{eq:codeword}).}
 \label{pic3}
\end{figure}

Shor was the first to overcome all these obstacles \cite{Shor:95a}. He gave the most straightforward quantum
generalization of the repetition code. Suppose we want to just deal with bit flip errors. We encode a single qubit with
the repetition code on the basis states, i.e.
 \be
 \ket{0} \longrightarrow \ket{000} \quad \quad \quad \ket{1} \longrightarrow \ket{111},\nonumber
 \ee
such that \be \label{eq:codeword}
 \alpha\ket{0}+\beta \ket{1} \longrightarrow \alpha \ket{000}+ \beta \ket{111}.
 \ee
This encoding can be realized with the circuit in Fig. \ref{pic3}.

\begin{figure}[h]
 \center{\epsfxsize=2.5in \epsfbox{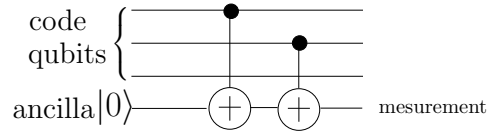}}
  \caption{\small Circuit to measure the parity of the first two qubits of the quantum code word. Each CNOT flips the
  ancilla qubit if the source qubit is in the state $\ket{1}$. If the first two qubits are in the state $\ket{00}$, the ancilla is left in
  the state $\ket{0}$. If these qubits are in the state $\ket{11}$ the ancilla is flipped twice and its state is also
  $\ket{0}$. Otherwise it is flipped once by one of the CNOTs.}
 \label{pic4}
\end{figure}

Now suppose a bit flip happens, say on the first qubit. The state becomes $\alpha \ket{100}+ \beta \ket{011}$. If we
measured the qubits in the computational basis, we would obtain one of the states $\ket{100}$ or $\ket{011}$, but we
would destroy the quantum superposition. But what if instead we measured the {\em parity} of all pairs of qubits,
without acquiring any additional information? For instance we can measure the parity of the first two qubits with the
circuit in Fig. \ref{pic4}.

In our example, a parity measurement does not destroy the superposition. If the first qubit is flipped, then both
$\ket{100}$ and $\ket{011}$ have the same parity $1$ on the first two qubits. If no qubit is flipped and the code word
is still in the state of Eq. (\ref{eq:codeword}) this parity will be $0$ for both $\ket{000}$ and $\ket{111}$. If the
error is a {\em linear combination} of identity and bit flip, similar to Eq. (\ref{eq:error}), then the measurement
will {\em collapse} the state into one of the two cases. Let us adapt Eq. ({\ref{eq:error}) to the case of only a bit
flip error on one qubit ($\ket{E_{00}}=\ket{E_{11}}$, $\ket{E_{01}}=\ket{E_{10}}$ and $\ket{E_{01}}$ and $\ket{E_{00}}$
are orthogonal) and write
 \begin{align}\label{eq:errorbf}
\left( \alpha\ket{0} + \beta \ket{1} \right) \otimes \ket{E} \quad  \longrightarrow \quad &
\sqrt{1-\eps}\underbrace{\left( \alpha \ket{0} + \beta \ket{1} \right)}_{identity} \otimes  \ket{\widetilde{E}_{00}}
 + \sqrt{\eps} \underbrace{\left( \alpha \ket{1} + \beta \ket{0} \right)}_{bit \, flip} \otimes
 \ket{\widetilde{E}_{01}},
 \end{align}
where we have normalized the state of the environment ($\ket{\widetilde{E}_{00}}$ and $\ket{\widetilde{E}_{01}}$ have
norm $1$)\footnote{Note, that if we trace out the environment (see App. \ref{sec:OSR}), we obtain a process where with
probability $1-\eps$ nothing happens, and with probability $\eps$ the bit is flipped. $\eps$ defines the {\em rate} of
error.}. The probability that the parity measurement collapses to the bit flip case is $\eps$, the probability to
project onto a state where no error has happened is $1-\eps$. Imagine now that each of the three qubits of the code
undergoes the same error process of Eq. (\ref{eq:errorbf}). This gives a threefold tensor product of Eq.
(\ref{eq:errorbf}) (each qubit has its own environment state), which shows that the probability of no error becomes
$(1-\eps)^3 \geq 1-3\eps$, and the probability of each of the single qubit errors is $\eps(1-\eps) < \eps$. Of course
there is now a nonzero probability that the state will be collapsed to a state where two or even three single qubit
errors occured; however, the total probability of this happening is given by $3 \eps^2(1-\eps)+\eps^3 \leq 3 \eps^2$.

This mechanism illustrates how a measurement that detects the error, also discretizes it. The parity measurement {\em
disentangles} the code qubits from the environment and acquires information about the error. The three parities (for
each qubit pair of the code word) give complete information about the location of the bit flip error. They constitute
what is called the error {\em syndrome} measurement. The syndrome measurement does not acquire any information about
the encoded superposition, and hence it does not destroy it. Depending on the outcome of the syndrome measurement, we
can correct the error by applying a bit flip to the appropriate qubit.

We have successfully resolved the introduction of redundancy, the discretization of errors and a way to measure the
syndrome without destroying the information. We still need to take care of phase flip errors. We have been able to
protect against bit flip errors by encoding the bits redundantly. The idea is to also encode the phase of the state in
a redundant fashion. Shor's idea was to encode a qubit using {\em nine} qubits in the following way:
 \begin{align}\label{eq:Shorcode}
& \ket{0}_{enc}=\frac{1}{\sqrt{2^3}} \left( \ket{000}+\ket{111} \right)\left( \ket{000}+\ket{111}
\right)\left( \ket{000}+\ket{111} \right) \nonumber \\
& \ket{1}_{enc}=\frac{1}{\sqrt{2^3}} \left( \ket{000}-\ket{111} \right)\left( \ket{000}-\ket{111} \right)\left(
\ket{000}-\ket{111} \right)
 \end{align}
Note that with this encoding, each of the blocks of three qubits is still encoded with a repetition code, so we can
still correct bit flip errors in a fashion very similar to above. But what about phase errors? A phase flip error, say
on one of the first three qubits, acts as:
\begin{align*}
&\ket{0}_{enc} {\stackrel{phase \, flip}{\longrightarrow}} \frac{1}{\sqrt{2^3}} \left( \ket{000}-\ket{111}
\right)\left( \ket{000}+\ket{111}
\right)\left( \ket{000}+\ket{111} \right) \nonumber \\
&\ket{1}_{enc} \stackrel{phase \, flip}{\longrightarrow} \frac{1}{\sqrt{2^3}} \left( \ket{000}+\ket{111} \right)\left(
\ket{000}-\ket{111} \right)\left( \ket{000}-\ket{111} \right)
 \end{align*}
We need to detect this phase flip {\em without} measuring the information in the state. To achieve this we will follow
the ideas developed for the bit flip and measure the parity of the phases on each pair of two of the three blocks.
There is an interesting and useful duality between bit flip and phase flip errors. Let us look at a different basis for
qubits, given by the states
 \be
  \ket{+}=\frac{1}{\sqrt{2}}\left(\ket{0}+\ket{1} \right) \quad \quad \ket{-}=\frac{1}{\sqrt{2}}\left(\ket{0}-\ket{1}
  \right)\nonumber
 \ee
The change from the standard basis to the $\ket{\pm}$-basis we apply the Hadamard transform $H$. Now note that a phase
flip error acts as
 \be \label{eq:bitphase}
  \ket{+} \stackrel{phase \, flip}{\longrightarrow} \ket{-} \quad \quad
  \ket{-} \stackrel{phase \, flip}{\longrightarrow} \ket{+}.
 \ee
In other words a phase flip in the standard basis becomes a {\em bit flip} in the $\ket{\pm}$-basis. If we apply a
Hadamard transform to each of the three qubits of a block of the Shor code, we obtain
 \begin{align*}
 H^{\otimes 3} \frac{1}{\sqrt{2}}\left(\ket{000}+\ket{111}\right) &=\frac{1}{2} \left(
 \ket{000}+\ket{110}+\ket{101}+\ket{011}\right)\nonumber \\
 H^{\otimes 3} \frac{1}{\sqrt{2}}\left(\ket{000}-\ket{111}\right) &=\frac{1}{2} \left(
 \ket{111}+\ket{001}+\ket{010}+\ket{100}\right)
 \end{align*}
Note that the parity of each of the bitstrings for positive phase is {\em even}
and for negative phase it is {\em odd}. We can see that if two blocks have
different phase, then  the parity of its constituent $6$ qubits is odd,
otherwise it is even. Hence, in order to detect a phase error, we just need to
measure the parity of all qubits in the three possible pairs of blocks in the
$\ket{\pm}$ basis. The circuit in Fig. \ref{pic5} does exactly that.

\begin{figure}[h]
 \center{\epsfxsize=3.5in \epsfbox{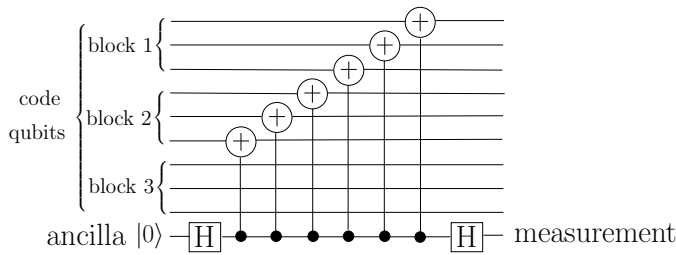}}
  \caption{\small Circuit to measure the parity of the {\em phase} of the first and the second block of three qubits.
  In the $\ket{\pm}$-basis a CNOT acts on target (t) and control (c) bit as $\ket{+}_t\ket{\pm}_c \rightarrow \ket{+}_t \ket{\pm}_c$ and
  $\ket{-}_t\ket{\pm}_c \rightarrow \ket{-}_t \ket{\mp}_c$, i.e. it flips the {\em control} bit in the $\ket{\pm}$ basis
  if the {\em target} bit is $\ket{-}_c$. This way the ancilla bit is flipped an even number of times from $\ket{+}$
  to $\ket{-}$ if blocks $1$ and $2$ have the same phase, and an odd number of times if they have different phase.}
 \label{pic5}
\end{figure}

The nine-bit Shor code above protects against bit and phase flip, and also against a combination of both (when {\em
both} bit and phase flip are detected, the error is $XZ$). Note again, that we assume that {\em each} of the qubits
undergoes some error at rate $\eps$. Hence, by the discretization resulting from the error-recovery measurement, the
state will be projected onto either a state where no error has occured (with probability $\geq 1- 9 \eps$) or a state
with a large error (single qubit, two qubit etc.). This code protects against all single qubit errors. Only when two or
more errors occur (which happens with probability $\leq 36 \eps^2$) the error is irrecoverable. Comparing this with the
error rate of an unencoded qubit, $\eps$, we see that this code is advantageous whenever $\eps \leq 1/36$.

\section{Quantum Error Correcting Codes}

 \begin{center}
 {\em ``If people do not believe that mathematics is simple, it is only because they do not realize how complicated life
is."\\ John von Neumann}
 \end{center}

Let us internalize the crucial properties of Shor's code: A small part of the Hilbert space of the system is designated
as the {\em code subspace} $\cal C$. In the Shor code $\cal C$ is spanned by the two states in Eq. (\ref{eq:Shorcode}).
We have a discrete set of correctable errors $\{\bf E_\alpha\}$. Each of the correctable errors ${\bf E_\alpha}$ maps
the code space $\mathcal C$ to a mutually {\em orthogonal} error space. We can make a measurement that tells us in
which of the mutually orthogonal spaces the system resides, and hence exactly infer the error. The error can be
repaired by applying an appropriate unitary transformation ($\bf E^\dagger_\alpha$).

These ideas have been formalized to define {\em quantum error correcting codes (QECCs)}. An $(N,K)$ quantum error
correcting code $\mathcal C$ is a $K$ dimensional subspace of an $N$ dimensional Hilbert space (coding space ${\mathcal
H}$) together with a recovery (super)operator $\mathcal R$. The recovery operator usually consists of some sort of
measurement (to detect the error) followed by a conditional unitary to correct it, but we do not necessarily have to
think about it in this way. The code $\mathcal C$ is {\em ${\mathcal E}$-correcting} if on the code-space an error
followed by recovery restores the codeword, i.e.
 \be {\mathcal R} \circ
{\mathcal E} = {\mathcal I} \quad \mbox{on} \quad {\mathcal C} \nonumber
 \ee
 It has been shown \cite{Bennett:96a,Knill:97b} that
QECC's exist for the set of errors if the following conditions ({\em QECC-conditions}) are satisfied:

\paragraph{\bf QECC-conditions:} Let ${\bf E}$ be a discrete linear base set for $\mathcal E$ and let the code $\mathcal C$ be
spanned by the basis $\{|\Psi_i\ra:\, i=1 \ldots K\}$. Then $\mathcal C$ is an $\mathcal E$-correcting QECC if and only
if  $\forall |\Psi_{i}\rangle, |\Psi_{j}\rangle \in {\mathcal C}$
\begin{equation}
\langle \Psi _{j}|{\bf E}_{\beta }^{\dagger }{\bf E}_{\alpha }|\Psi _{i}\rangle =c_{\alpha \beta }\delta _{ij}\quad
\forall {\bf E_{\alpha },E_{\beta }}\in {\bf E}.  \label{eq:QECCcond1}
\end{equation}
What this means is the following: Errors ${\bf E_{\alpha },E_{\beta }}\in {\bf E}$ acting on {\em different} orthogonal
codewords $|\Psi_{i}\rangle$ take these codewords to {\em orthogonal} states ($\langle \Psi _{i}|{\bf E}_{\beta
}^{\dagger }{\bf E}_{\alpha }|\Psi_{j}\rangle =0$). Otherwise errors would destroy the perfect distinguishability of
orthogonal codewords and no recovery would be possible. On the other hand for different errors acting on the {\em same}
codeword $|\Psi_i\ra$ we only require that $\langle \Psi _{i}|{\bf E}_{\beta }^{\dagger }{\bf E}_{\alpha
}|\Psi_{i}\rangle$ does not depend on $i$. Otherwise we would - in identifying the error - acquire some information
about the encoded state $|\Psi_i\ra$ and thus inevitably disturb it.

We usually think of the errors $\bf E_\alpha$ to be a subset of the Pauli group with up to $t$ non-identity Pauli
matrices (for a $t$-error correcting QECC).

It is now possible to make the connection to the theory of classical error
correcting codes. It turns out that there are families of classical codes with
certain properties (concerning their dual) which make good quantum error
correcting codes \cite{Steane:96a,Calderbank:96a}. The codes have become known
as Calderbank-Shor-Steane codes (CSS codes). It has been shown that for any
number $t$ of correctable errors, there is a QECC which can correct up to $t$
errors (bit flip, phase flip and combination). As a result this code reduces
the error  for an unencoded qubit, $\eps$, to $c \eps ^{t+1}$, where $c$ is a
constant depending on the code.

To illustrate this connection to classical codes we will briefly describe the smallest code in that family, which was
first given by Steane \cite{Steane:96a}.  This is the so called $7$ qubit {\em Steane code}, based on the classical
$7$-bit Hamming code. The classical Hamming code encodes one bit into 7 bits. The codewords can be characterized by the
{\em parity check} matrix
\begin{equation}\label{eq:Hamming}
H=\left( \begin{array}{ccccccc}0&0&0&1&1&1&1\\
    0&1&1&0&0&1&1\\
    1&0&1&0&1&0&1\end{array}\right).
\end{equation}
The code is the kernel of $H$, i.e. each code word is a 7-bit vector $v_{code}$ such that $H \cdot v_{code}=(0,0,0)^T$
in $GF(2)$ arithmetic. $H$ has three linearly independent rows (over $GF(2)$), so the kernel is spanned by four
linearly independent code words, and hence there are $16$ different code words. If an error affects the $i$th bit of
the codeword, this codeword is changed to $v_{code}+e_i$. The parity check matrix of the resulting word is
$H(v_{code}+e_i)=He_i \neq 0$, which is just the $i$th column of $H$. Since all columns of $H$ are distinct, each $e_i$
has a different {\em error syndrome} and we can infer $e_i$ from it.

Steane's code, derived from the Hamming code, is the following
\begin{eqnarray} \label{eq:Steane}
\label{zero} & |0\rangle_{code}={1\over\sqrt{8}} \left(\sum_{{\rm even}~v\atop \in ~{\rm Hamming}}|v\rangle\right)
={1\over \sqrt{8}} & \Big(  |0000000\rangle + |0001111\rangle +|0110011\rangle
+|0111100\rangle\nonumber\\
 && +  |1010101\rangle  +|1011010\rangle +|1100110\rangle +|1101001\rangle\Big)
\ ,\nonumber\\
& |1\rangle_{code}={1\over\sqrt{8}}\left(\sum_{{\rm odd}~v\atop \in ~{\rm Hamming}}|v\rangle \right) ={1\over \sqrt{8}}
&\Big(   |1111111\rangle  + |1110000\rangle +|1001100\rangle
+|1000011\rangle\nonumber\\
&& + |0101010\rangle  +|0100101\rangle +|0011001\rangle +|0010110\rangle\Big),\nonumber \\
\end{eqnarray}
i.e. $\ket{0}_{code}$ is the superposition of all even and $\ket{1}_{code}$ the superposition of all odd codewords.
Note that all states appearing in the code words are Hamming code words, and hence a single bit flip can be detected by
a simple parity measurement, as in Fig. \ref{pic6}.

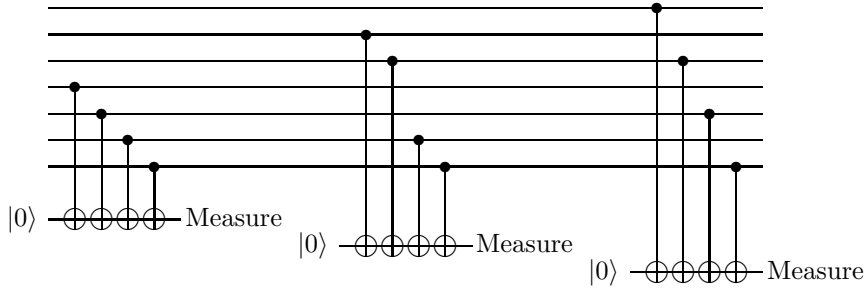
\begin{figure}
\centering
\begin{picture}(320,115)

\put(0,30){\makebox(0,0){$\ket{0}$}} \put(70,25){\makebox(20,12){Measure}}

\put(10,110){\line(1,0){270}} \put(10,100){\line(1,0){270}} \put(10,90){\line(1,0){270}} \put(10,80){\line(1,0){270}}
\put(10,70){\line(1,0){270}} \put(10,60){\line(1,0){270}} \put(10,50){\line(1,0){270}}

\put(10,30){\line(1,0){50}}

\put(20,80){\circle*{4}} \put(20,80){\line(0,-1){54}} \put(20,30){\circle{8}}

\put(30,70){\circle*{4}} \put(30,70){\line(0,-1){44}} \put(30,30){\circle{8}}

\put(40,60){\circle*{4}} \put(40,60){\line(0,-1){34}} \put(40,30){\circle{8}}

\put(50,50){\circle*{4}} \put(50,50){\line(0,-1){24}} \put(50,30){\circle{8}}

\put(110,20){\makebox(0,0){$\ket{0}$}} \put(180,15){\makebox(20,12){Measure}}

\put(120,20){\line(1,0){50}}

\put(130,100){\circle*{4}} \put(130,100){\line(0,-1){84}} \put(130,20){\circle{8}}

\put(140,90){\circle*{4}} \put(140,90){\line(0,-1){74}} \put(140,20){\circle{8}}

\put(150,60){\circle*{4}} \put(150,60){\line(0,-1){44}} \put(150,20){\circle{8}}

\put(160,50){\circle*{4}} \put(160,50){\line(0,-1){34}} \put(160,20){\circle{8}}

\put(220,10){\makebox(0,0){$\ket{0}$}} \put(290,5){\makebox(20,12){Measure}}

\put(230,10){\line(1,0){50}}

\put(240,110){\circle*{4}} \put(240,110){\line(0,-1){104}} \put(240,10){\circle{8}}

\put(250,90){\circle*{4}} \put(250,90){\line(0,-1){84}} \put(250,10){\circle{8}}

\put(260,70){\circle*{4}} \put(260,70){\line(0,-1){64}} \put(260,10){\circle{8}}

\put(270,50){\circle*{4}} \put(270,50){\line(0,-1){44}} \put(270,10){\circle{8}}

\end{picture}
\caption{Computation of the bit-flip syndrome for Steane's 7-qubit code. The three ancilla qubits carry the error
syndrome.} \label{pic6}
\end{figure}
To deal with phase flip errors we use the observation of Eq. (\ref{eq:bitphase}), that phase flip errors correspond to
bit flip errors in the $\ket{\pm}$ basis. But if we change to this basis by applying the Hadamard transform to each
bit, we obtain

\begin{eqnarray}
\label{eq:steanehadamard}
 H^{\otimes 7}| 0\rangle_{\rm code}=&{1\over 4}\left(\sum_{v \in \atop {\rm Hamming}}|v\rangle\right) ={1\over
\sqrt{2}}\left(|0\rangle_{\rm code} + |1\rangle_{\rm code}\right) \
,\nonumber\\
 H^{\otimes 7}|1\rangle_{\rm code}=&{1\over 4}\left(\sum_{v \in \atop {\rm Hamming}}(-1)^{wt(v)}|v\rangle \right) ={1\over
\sqrt{2}}\left(|0\rangle_{\rm code} - |1\rangle_{\rm code}\right),
\end{eqnarray}
(where $wt(v)$ denotes the weight of $v$). The key point is that in the $\ket{\pm}$ basis, like in the
$\ket{0},\ket{1}$ basis, $|0\rangle_{code}$ and $|1\rangle_{code}$, are superpositions of Hamming codewords. Hence, in
the rotated basis, as in the original basis, we can perform the Hamming parity check to diagnose bit flips, which are
phase flips in the original basis. Assuming that only one qubit is in error, performing the parity check in both bases
completely diagnoses the error, and enables us to correct it.

The core observation that allows to generalise Steane's construction to codes that encode more bits and can correct
more errors is the following: If a quantum code word is a linear superposition over classical code words that form a
code $\mathcal C$, then in the $\ket{\pm}$ basis this code word is a linear superposition over the code words of the
{\em dual} code ${\mathcal C}^\perp$, where $ {\mathcal C}^\perp=\{u:\, u \cdot v=0 \,\, \forall v \in {\mathcal C}\}$.
This can derived when looking at the action of the Hadamard transform on $n$-bit strings $\ket{x}$:
 \be
 H^{\otimes n} \ket{x} = \frac{1}{\sqrt{2^n}} \sum_{y \in \{0,1\}^n} (-1)^{x \cdot y} \ket{y}\nonumber
 \ee
 As it is easy to see from Eq.
(\ref{eq:Hamming}), the Hamming code is its own dual, and hence we can use its properties to correct phase errors. In
general, the CSS constructions find a code ${\mathcal C}_1$ (for the bit flip errors) such that its dual, ${\mathcal
C}_1^\perp$, contains a sufficiently good code ${\mathcal C}_2$ (for the phase flip errors).

Having seen the nine qubit Shor code and the seven qubit Steane code, one can ask what the minimal overhead for a
quantum code that corrects a single error is. It turns out that the smallest quantum code that achieves this has five
qubits, and that this is optimal \cite{Laflamme:96a}.

Gottesman developed a very powerful formalism, so called {\em stabilizer codes}, that generalises both the Shor code
and CSS codes and gave fault tolerant constructions for them (for more details see App. \ref{sec:stabilizer}).

\section{Fault-tolerant computation}
\begin{center}
{\em ``When you have faults, do not fear to abandon them."\\Confucius}
\end{center}

We have seen that good quantum error correction codes exist. But so far we have worked under the assumption, that the
error recovery procedure is {\em perfect}. Of course, error recovery will never be flawless. Recovery is itself a
quantum computation that will be prone to decoherence. We must ensure that errors do not {\em propagate} during
recovery. For instance, if an error occurs in the ancilla bit in the parity measurement of Fig. \ref{pic5}, {\em all}
six qubits interacting with it might be corrupted; the error propagates catastrophically. In fault-tolerant computing
design, care is taken to avoid this type of error spreading, and other possible introduction and propagation of error.
But even if we manage to avoid error spreading during recovery, that is not enough. A quantum computer does more than
just {\em store} information, it also {\em processes} it. Of course we could decode, perform a gate and encode, but
this procedure would temporarily expose quantum information to decoherence. Instead, we must apply our quantum gates
directly to the encoded data.

\subsection{Guidelines of fault-tolerance}

The quantum circuit model gives us a good intuition about the points in a computation that potentially can introduce
errors and corrupt the computation. We need to be able to faultlessly {\em prepare} the initial state, {\em compute}
with a sequence of quantum gates and {\em measure} the output. Using a code to protect our computation against noise,
we also need to assure faultless {\em encoding}, {\em decoding} and {\em correction}. Each qubit will be encoded into a
separate block and quantum logic has to be applied directly {\em on the encoded states} so that the information is
never exposed to noise without protection. This gives us the following guidelines of fault-tolerance:
\paragraph {\em Encoding/Decoding/State Preparation} The procedure to encode/decode the information into a code
should not introduce more errors than the code can correct. In the case of a 1-error correcting QECC encoding should
not introduce more than one error per encoded block. Often the only states that need to be encoded are some
$|00\ldots0\ra$ states at the beginning of the computation, it is then sufficient to ensure fault-tolerant {\em
state-preparation}.

\paragraph {\em Error-detection and Recovery} These procedures (for a QECC), usually realized by a set of quantum
gates together with auxiliary qubits, should again not introduce more than one error per block.

\paragraph{\em  Quantum gates} should not introduce more than one error per encoded block. Furthermore they should
not {\em propagate} already existing errors from one qubit to several others in the same block.

\paragraph{\em Measurement} should not introduce more than one error per block. Furthermore the measurement result
must have probability of error of order $\eps^2$, where $\eps$ is the probability of failure of any of the components
in the measurement procedure. This is because the measurement result may be used to control other operations in a
quantum computer.

\subsection{Fault-tolerant error correction}

With these guidelines in place, we will now illustrate how fault-tolerant recovery can be achieved, using the Steane
code in Eq.(\ref{eq:Steane}) as our example. The error-measurement circuit in Fig. \ref{pic6} is not fault-tolerant, as
each of the CNOT gates can propagate a single phase error on the ancilla qubit to all four of the code qubits. To
prevent this propagation, we need to expand the ancilla into four qubits, each one the target of only one CNOT gate.
But now we are again faced with the problem that our measurement should only reveal information about the {\em error}
(the parity) but not about the encoded state. We circumvent this problem by preparing the ancilla in the following
state
 \be
\ket{ancilla}=\frac{1}{\sqrt{8}}
\left(\ket{0000}+\ket{1100}+\ket{1010}+\ket{1001}+\ket{0110}+\ket{0101}+\ket{0011}+\ket{1111} \right),\nonumber
 \ee
i.e. in a superposition of all even bit strings. The crucial observation is that  on this state one bit flip or three
bit flips on any qubits all have the same effect, they transform it to the superposition of odd bit strings. Similarly,
this state is invariant under any even number of bit flips. This means that we can infer the syndrome bit from the
parity of the ancilla bits, it suffices to measure the ancilla in the end. Hence our syndrome measurement obeys the
guidelines of fault-tolerance. To prepare the ancilla, we can use the circuit in Fig. \ref{pic7}, which at the same
time allows {\em verification} of correct ancilla preparation.

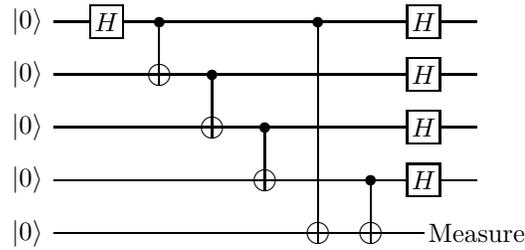
\begin{figure}
\centering
\begin{picture}(210,120)

\put(0,14){\makebox(20,12){$\ket{0}$}} \put(0,34){\makebox(20,12){$\ket{0}$}} \put(0,54){\makebox(20,12){$\ket{0}$}}
\put(0,74){\makebox(20,12){$\ket{0}$}} \put(0,94){\makebox(20,12){$\ket{0}$}}

\put(20,20){\line(1,0){140}} \put(20,40){\line(1,0){134}} \put(20,60){\line(1,0){134}} \put(20,80){\line(1,0){134}}
\put(20,100){\line(1,0){14}} \put(46,100){\line(1,0){108}}

\put(34,94){\framebox(12,12){$H$}}

\put(60,100){\circle*{4}} \put(60,100){\line(0,-1){24}} \put(60,80){\circle{8}}

\put(80,80){\circle*{4}} \put(80,80){\line(0,-1){24}} \put(80,60){\circle{8}}

\put(100,60){\circle*{4}} \put(100,60){\line(0,-1){24}} \put(100,40){\circle{8}}

\put(120,100){\circle*{4}} \put(120,100){\line(0,-1){84}} \put(120,20){\circle{8}}

\put(140,40){\circle*{4}} \put(140,40){\line(0,-1){24}} \put(140,20){\circle{8}}

\put(154,34){\framebox(12,12){$H$}} \put(154,54){\framebox(12,12){$H$}} \put(154,74){\framebox(12,12){$H$}}
\put(154,94){\framebox(12,12){$H$}} \put(160,14){\makebox(40,12){Measure}}

\put(166,40){\line(1,0){14}} \put(166,60){\line(1,0){14}} \put(166,80){\line(1,0){14}} \put(166,100){\line(1,0){14}}

\end{picture}
\caption{Construction and verification of the $\ket{ancilla}$ state.  If the measurement outcome is 1, then the state
is discarded and a new $\ket{ancilla}$ state is prepared.} \label{pic7}
\end{figure}
The ancilla state must be verified before it is used, because a single error in the preparation of the ancilla state
can propagate and cause two phase errors in the $\ket{ancilla}$ state. Hence the circuit in Fig. \ref{pic7} also
verifies that multiple phase errors do not occur. If it fails the test it should be discarded, and the preparation
procedure repeated.

Moreover, a single syndrome measurement might be faulty. Thus, the syndrome measurement should be repeated for
accuracy; only if the same result is measured twice in a row should it be accepted.

With all the precautions above, recovery will only fail if two independent errors occur in this entire procedure. The
probability that this happens is still  $c\eps^2$ for some constant $c$, but because there are now many more gates and
steps involved the constant $c$ can be quite large.

In a conceptually similar fashion it is possible to {\em encode} a qubit and to measure it in a basis spanned by
$\ket{0}_{code}$ and $\ket{1}_{code}$ while following the guideline of fault-tolerance. For details the reader should
consult e.g. \cite{Shor:96a,Steane:97a,Preskill:98a,Preskill:99a,Preskill:notes} or the work of Gottesman (e.g.
\cite{Gottesman:97b}) for fault-tolerant construction for CSS and other codes in the stabilizer formalism (see App.
\ref{sec:stabilizer}).

\subsection{Fault-tolerant computation}

We have seen how to recover {\em stored} quantum information, even when
recovery is faulty. But we also want to {\em compute}, and the gates we use
will be faulty as well. This means that we must be able to apply the gates
directly to the encoded data, without introducing errors uncontrollably;
following the guidelines of fault-tolerance.

In fact, staying with the $7$-qubit Steane code, it is easy to implement some single qubit gates directly on the
encoded data. We have seen that the bitwise Hadamard transform implements an {\em encoded} Hadamard transform on the
codewords (see Eq. (\ref{eq:steanehadamard})). This means we can apply it without propagating errors and such that each
gate introduces at most one new error. Similarly, it is easy to see that the bitwise $X$ gate induces an encoded
$X_{enc}$ because even code words get mapped to odd ones and vice versa. Moreover the bitwise $Z$ gate (which is just
$HXH$) implements the encoded $Z$. In the same way the $\frac{\pi}{4}$ gate (a diagonal single qubit unitary with
diagonal $(1,i)$) can be implemented by applying it bitwise to the encoded data.

Also, it is not hard to see that the bitwise CNOT between two quantum code words, i.e. a CNOT from the first qubit of
the first code word to the first qubit of the second code word, a CNOT from the second qubit of the first code word, to
the second qubit of the second and so on, implements a global CNOT between two code words. We call such an
implementation of an encoded two qubit gate {\em transversal}. This is very promising, but the set of operation we can
implement fault-tolerantly is not yet universal. We also need to implement the $\pi/8$ gate for a universal set of
gates. Unfortunately it seems to be impossible to implement the $\pi/8$ gate in a fault-tolerant way. There are several
ways to circumvent this problem. Shor, for instance, gave a way to complete the universal set by giving a transversal
implementation of a three qubit gate, the Toffoli gate \cite{Shor:96a}. However, we will follow a slightly different
route here. It turns out that the gates $\{X,\frac{\pi}{4},CNOT\}$ are universal, provided we can measure a code word
in the $\ket{0}_{code}$, $\ket{1}_{code}$ basis and we have access to the state
 \be
 \ket{\pi/8_+}_{code}=\ket{0}_{code}+ \exp(i \frac{\pi}{4}) \ket{1}_{code}.\nonumber
 \ee
It has been shown that there is a fault-tolerant preparation and verification procedure for the state
$\ket{\pi/8_+}_{code}$, which is similar in spirit to the one in Fig. \ref{pic7}.

Several variants of fault-tolerant universal quantum computation have been developed for this and other codes, like CSS
codes and stabilizer codes. They differ in the details of ancilla preparation and number of interactions with the code
word. As a result it is possible to implement computation and error correction following the guidelines of
fault-tolerance.

\section{Concatenated coding and the threshold}\label{sec:threshold}
\begin{center}
{\em ``Much of modern art is devoted to lowering the threshold of what is terrible."\\Susan Sontag}
 \end{center}

We have seen how to encode quantum data, how to perform fault-tolerant recovery and how to compute fault tolerantly on
encoded states. However, this is still not sufficient to implement quantum algorithms. Quantum codes exist that can
correct up to $t$ errors, where $t$ can be as large as we wish, and on which we can compute fault-tolerantly. This
means that if our error rate and gate and measurement failure rate is $\eps$, then computation will only fail with
probability of order $\eps^{t+1}$ for a $t$ of our choice. So what is the problem?

The crux is the complexity of the recovery procedure. With large $t$ we reach a point where the recovery procedure
takes so much time that it becomes likely that $t+1$ errors occur in a block. The number of steps required for recovery
scales as a power of $t$, $t^a$ with exponent $a>1$. That means that the probability to have $t+1$ errors before a
recovery step is completed, scales as $(t^a \eps)^{t+1}$. This expression is minimized when $t=c \eps^{-\frac{1}{a}}$
for some constant $c$ and its value is at least $p_{fail}=\exp(-ca\eps^{-\frac{1}{a}})$. This means that per error
correction cycle our probability to fail is at least $p_{fail}$. If we have $N$ such cycles, our failure probability is
$N p_{fail}=\exp(-ca \log N \eps^{-\frac{1}{a}})$. If we want to keep this (much) smaller than $1$, our error rate
$\eps$ has to scale as $\frac{1}{(\log N)^a}$, i.e. the longer the computation, the more accuracy we need; an
unrealistic assumption.

To overcome this problem, a special kind of hierarchical approach is used \cite{Knill:98a} (see also
\cite{Kitaev:97a,Aharonov:97a}). Ideas related to this approach go back to pioneering works of John von Neumann, who
established a theory of fault-tolerant computation for noisy classical computers \cite{vonNeumann:56a}.

 Suppose that we encode our information into a code, like Steane's code. Then, in turn,
we encode each qubit of this encoded qubit using again Steane's code, and so on. We obtain several layers of encoded
qubits, say $k$ layers, and the total number of qubits is $7^k$. This type of code is called {\em concatenated} code.

The exact calculations behind the threshold theorem are rather intricate. Let us only give a rough intuition. The idea
is to perform error recovery most often at the lowest level, and less and less often at higher levels of the hierarchy,
which have more qubits. We recursively apply the idea of simulating a circuit using an encoded circuit, constructing a
hierarchy of quantum circuits. Suppose in the first stage the original qubit is encoded in a quantum code whose encoded
qubits are again encoded in a quantum code and so on. Each level has some error recovery cycles. If the failure
probability at the lowest level of this code is $\eps$ then the failure probability at the next level of encoding is
$c\eps^2$ (remember that the Steane code reduces the error rate from first to second order), where $c$ counts all
possibilities that two errors can occur, given the number of gates in the recovery procedure and the fault-tolerant
application of gates. Continuing with this reasoning, the {\em effective} error rate at the next level is $c \eps^2$,
and error recovery reduces the error to $c(c\eps^2)^2$. Proceeding level by level, we see that at the $k$th level of
the hierarchy an error on one of the sub-blocks only has probability $(c\eps)^{2^k}/c$. We see that if our noise rate
is below a certain threshold, $\eps < \eps_{th}\equiv 1/c$, then the error is reduced in each level of concatenation.
This gives the {\em error threshold} for fault-tolerant quantum computation.

How does the total size of the circuit grow? Let's assume that one level of encoding requires an overhead of $G$ gates
to fault-tolerantly perform a gate and error-correct. Then the size of the simulating circuit grows as $G^k$. Let us
see when this concatenation procedure gives a small enough failure probability:

Assume the initial quantum circuit we want to emulate has $N$ gates and we wish to achieve final accuracy $1-p$. In
such a circuit each gate has to have a failure probability less than $p/N$ (gate errors add linearly). To achieve this
we concatenate $k$ times so that \be \frac{(c \eps)^{2^k}}{c}=\eps_{th}(\frac{\eps}{\eps_{th}})^{2^k} \leq
\frac{p}{N}\nonumber
 \ee
 or
 \be
2^k \leq \frac{\log (N \eps_{th}/ p)}{\log (\eps_{th}/\eps)}\nonumber
 \ee
If $\eps$ is smaller than the threshold value, such a $k$ can be found. For error rates below the threshold we can
achieve arbitrary accuracy by concatenation. Per initial gate the final circuit will have
 \be G^k=2^{k\log G} \leq \left(\frac {\log (N \eps_{th}/ p)}{\log(\eps_{th}/\eps)}\right)^{\log G}=
 poly(\log N) \nonumber \ee
 gates and so its final size will be $Npoly(\log
N)$ which is only polylogarithmically larger than the original $N$.

Note that we have crudely simplified our calculations. Estimating the threshold is an extremely intricate task. Its
value depends on the details of the code and fault-tolerance constructions used. It also depends on whether we assume
the classical syndrome processing to be perfect or not. In all cases it seems that we need high parallelization and a
supply of fresh ancilla qubits during the computation. For a long time the actual value of the threshold has  been
estimated by optimists and pessimists to lie somewhere between $10^{-4}$ and $10^{-7}$. Recent work seems to indicate
that it can be even as high as $3\%$ \cite{Knill:05a} (see also \cite{Aliferis:05a,Reichardt:05a}) and optimized
numerical simulations of fault-tolerant protocols suggest a threshold as high as $5\%$ (however, to tolerate this much
error existing protocols require enormous overhead).

\section{Error avoidance and Decoherence Free Subsystems}
\begin{center}
{\em ``It is well known that ``problem avoidance" is an important part of problem solving."\\ Edward de Bono}
\end{center}

In all our previous analysis we have assumed that the errors behave independently and affect few qubits at a time.
What, if this is not the case? There are situations, where groups of qubits interact with the environment in a {\em
collective} fashion, possibly undergoing a correlated error. For these cases the theory of decoherence-free subspaces
and subsystems has been developed, sometimes also called {\em error-avoiding} codes. These codes come into play when
the decoherence process is in some sense not local, but collective, involving groups of qubits.

Let us give a classical example. Assume we have an error process that with some probability flips all bits in a group,
and otherwise does nothing. In this case we can encode a classical bit as
 \be
 0 \longrightarrow 00 \quad \quad \quad 1 \longrightarrow 01.\nonumber
 \ee
The error process will change the encoded states to \be
 00 \longrightarrow 11 \quad \quad \quad 01 \longrightarrow 10.\nonumber
 \ee
But no matter if the error has acted or not, the {\em parity} of the bit string is unchanged. So when we decode, we
will associate $00$ and $11$ with the encoded $0$ bit and $01$ and $10$ with the encoded $1$. Note that we will be able
to decode correctly {\em no matter} how hight the rate of error is! The error does not touch the invariant, parity,
into which we encode. That means that our encoded information has managed to completely {\em avoid} the error, we have
given the simplest error-avoiding code.

A lot of research has been done to generalize this to the quantum case (see e.g. \cite{Kempe:PHD,Bacon:PHD,Lidar:03a}
for surveys). The noise model is in general derived from the Hamiltonian picture (see App. \ref{sec:Hamiltonian}) or
from the Markovian picture (see App. \ref{sec:markovian}), a brief derivation is given in App. \ref{sec:DFS}. In
general the underlying assumption is that several qubits couple collectively to the environment and are affected by a
symmetric decoherence process. In systems where this form of decoherence is dominant at the qubit level, error-avoiding
codes as part of the error-correction scheme are advantageous.

We will content ourselves with briefly describing one example. For one of the most common collective decoherence
processes the noise operators on $n$ qubits (see App. \ref{sec:DFS}) in the Hamiltonian picture (App.
\ref{sec:Hamiltonian}) are given by $S_\alpha=\sum_{i=1}^n \sigma_\alpha^i$, where $\sigma^i_\alpha$ is a Pauli matrix
($\alpha = \{x,y,z\}$) on the $i$th qubit. Intuitively this means that the possible unitary errors are
$\exp(itS_\alpha)$. The condition for decoherence-free subspaces is that $$S_{\alpha}\ket{codeword}=c_\alpha
\ket{codeword},$$ or in other words that the code space is a simultaneous eigenspace of each $S_\alpha$ with eigenvalue
$c_\alpha$. If this is the case, each unitary noise operator only introduces an unobservable phase $exp(i t c_\alpha)$
on the code space.

Let us look at an encoding of $4$ qubits:
\begin{eqnarray*}
|0\rangle_{code}  &=&|s\rangle \otimes |s\rangle   \nonumber \\
|1\rangle_{code}  &=&\frac{1}{\sqrt{3}}\left(|t_{+}\rangle \otimes |t_{-}\rangle -|t_{0}\rangle \otimes |t_{0}\rangle
+|t_{-}\rangle \otimes |t_{+}\rangle \right),
\end{eqnarray*}
where $|s\rangle =\frac{|01\rangle -|10\rangle }{\sqrt{2}}$  and $|t_{-,0,+}\rangle =\{|00\rangle ,\frac{|01\rangle
+|10\rangle }{\sqrt{2}},|11\rangle \}$. It is easy to see that $S_\alpha \ket{0}_{code}=S_\alpha \ket{1}_{code}=0$ for
$\alpha=\{x,y,z\}$ (i.e. that the coefficients $c_\alpha=0$). This means that both code states are invariant under
collective noise. If we encode our information into the subspace spanned by $\ket{0}_{code}$ and $\ket{1}_{code}$, it
will completely avoid the errors; it resides in a ``quiet" part of the space, a {\em decoherence-free} subspace.

This idea has been generalized to a wide variety of encodings (in particular to decoherence-free {\em subsystems}, see
App. \ref{sec:DFS} for a little more detail) and against several kinds of collective noise. It has been shown how to
{\em compute} on these codes (e.g. \cite{Kempe:01a}) and how to use them in a fault-tolerant framework.

Of course the noise in a real implementation of a quantum computer will be a mixture of independent statistical noise,
and coupled collective noise, depending on the specific quantum hardware used. The idea is to use a hierarchical
construction of concatenated quantum codes, as in the threshold construction of Sec. \ref{sec:threshold}, where the
lower levels of the hierarchy use error correction (or avoidance) schemes that are highly specialized to the
anticipated noise process, whereas higher levels are similar to the known fault-tolerant constructions for QECCs (see
e.g. \cite{Lidar:99b} for a DFS-QECC concatenation scheme).

\section{Conclusion and Epilogue}
\begin{center}
{\em ``The best thing about the future is that it only comes one day at a time."\\Abraham Lincoln}
\end{center}

Without doubt work on quantum fault-tolerance is of prime importance if we want to build quantum machines. In the late
1990's pioneering work has established that fault-tolerant quantum computation is {\em possible}, and we have estimates
for the {\em error threshold}, the maximum error a component can undergo such that the computation still proceeds
without catastrophic error. On the way we have gained new insights into the nature of decoherence and about the methods
and tools used to model and describe it.

At this point we need to optimize the details of fault-tolerant schemes and to generate new ideas to improve the
threshold. Current experimental results show that the accuracy in implementations of the quantum circuit model is on
the order of several percent in the best case, whereas most estimates of the threshold give numbers of the order of
$10^{-4}$ or less.

The task for current research is to analyse the threshold for particular codes and to develop new elements of
fault-tolerance that improve the threshold. For instance only recently the existance of a threshold for the Steane code
has been shown \cite{Aliferis:05a, Reichardt:05a}. New elements have been developed to improve the threshold, like for
instance schemes based on postselection \cite{Knill:04a}. Using new ingredients, the threshold has now been estimated
to be on the order of $3\%$, albeit with an enormous overhead in the circuit architecture  \cite{Knill:05a}.

Another avenue of research is to explore other models of quantum computing, different from the quantum circuit model,
which can be inherently more robust against noise. One idea, initiated by Kitaev \cite{Kitaev:03a}, is a scheme for
intrinsically fault-tolerant quantum hardware, designed to be robust against localized inaccuracies. In this scheme
(quantum computing by anyons) gates exploit non-Abelian Aharonov-Bohm interactions among separated quasiparticles on a
2D lattice (see also e.g. \cite{Simon:05a}).

Another potentially more robust model is the model of adiabatic quantum computation, where computation is achieved by
adiabatically tuning a set of Hamiltonians, and where the system is always in the instantaneous groundstate. In this
system there is a gap between the groundstate and the first exited state at all times, which might make the state more
robust to noise (see \cite{Farhi:01a,Aharonov:04a,Childs:01b}). Yet another model is the measurement based {\em
one-way} quantum computer \cite{Briegel:01a}. Here quantum computation is achieved by measuring single qubits of a
suitably prepared initial state. The fault-tolerance properties of this system have recently been explored in
\cite{Nielsen:05a}.

All these developments allow us to be optimistic about the future of a quantum computing machine. One day we might be
able to combat decoherence and have large scale entangled states operate for us. The consequences not only for
computation, but also for our understanding of the fundamental processes behind decoherence will be formidable.

\appendix

\section{Further Reading}\label{sec:reading}

An ever growing community of researchers has been and is working an error correction and fault-tolerance in the quantum
setting, and it is impossible to mention all of them in this framework. What follows is a selection of some of the
milestones and recent developments, where the interested reader can find more information.

That quantum error correcting codes exist was first pointed out by Shor \cite{Shor:95a} and Steane \cite{Steane:96a} in
the end of 1995. By early 1996 it was shown by Steane \cite{Steane:96c} and Calderbank and Shor \cite{Calderbank:96a}
that {\em good} codes exist, i.e. codes that are capable to correct many errors. The quantum error correction
conditions where formalized by Knill and Laflamme \cite{Knill:97b} and Bennett et al. \cite{Bennett:96a}\footnote{These
authors also analyzed schemes based on random codes.}.

The first fully {\em fault-tolerant recovery} scheme, which takes into account that encoding, error-correction and
decoding are themselves noisy operations, was developed by Shor in 1996 \cite{Shor:96a}. Methods for fault-tolerant
recovery where independently developed by Kitaev.

The first to show that there is an {\em accuracy threshold} for {\em storage} of quantum information where Knill and
Laflamme \cite{Knill:96a} and for {\em quantum computation} Knill, Laflamme and Zurek \cite{Knill:96c} in 1996 (see
also \cite{Knill:98a}). Similar results were reported by Kitaev \cite{Kitaev:97a} and Aharonov and Ben-Or
\cite{Aharonov:97a}.

The theory of {\em stabilizer codes} and of fault-tolerance in the powerful stabilizer formalism was developed by
Gottesman (see e.g. \cite{Gottesman:97b})\footnote{See also work by Calderbank, Rains, Shor and Sloane, which develop
stabilizer codes as codes over GF(4) \cite{Calderbank:98a}}.

Since then several researchers have sharpened the threshold and developed new techniques to analyse and improve it. See
for instance the recent work of Steane \cite{Steane:99a, Steane:03a}, Knill \cite{Knill:04a,Knill:05a}, Aliferis,
Gottesman and Preskill \cite{Aliferis:05a} and  Reichardt \cite{Reichardt:05a} and \cite{Fern:04a} for a dynamical
systems approach.

In the literature the terms "sub-" and "superradiance" are often encountered in connection with {\em collective
decoherence} processes. Decoherence-free subspaces  have been studied by several researchers (see for example
\cite{Duan:98a, Zanardi:97a,Zanardi:97b,Zanardi:98a} in the context of storage, and
\cite{Lidar:98a,Bacon:99a,Bacon:99b,Kempe:01a,DiVincenzo:00a} in the context of fault-tolerant computation and
\cite{Lidar:99b} in combination with QECCs). Decoherence-free subsystems have been introduced by
\cite{Viola:99a,Knill:00a}, also in the connection with dynamic decoupling techniques. Since then there has been a lot
of active effort to adapt codes to various collective noise processes.

The threshold has also been inspected in the light of various (local) error models. For instance \cite{Terhal:05a}
discuss the fault-tolerant threshold for local {\em non-Markovian} noise (see also \cite{Alicki:05a} and references
therein for a recent controversy about the nature of errors and an analysis of error models that seem to not allow
fault-tolerant computation in \cite{Kalai:05a}).

Apart from the quantum circuit model alternative proposals for quantum architectures have been developed, which could
be potentially more robust than the quantum circuit model. One example, developed by Kitaev, is the model of
computation via {\em anyons} (see \cite{Kitaev:03a} for an analysis of its fault-tolerance properties, or e.g. Freedman
et al. \cite{Freedman:01a}). Another recent example is the measurement based {\em quantum cluster} model
\cite{Briegel:01a} introduced by Briegel and Raussendorf. Recently, fault-tolerance has been analyzed in this model by
Nielsen and Dawson \cite{Nielsen:05a}. The {\em adiabatic model} of quantum computation has been introduced by Farhi et
al. \cite{Farhi:01a} and shown to be equivalent to the quantum circuit model in \cite{Aharonov:04a}. Childs et al. have
discussed its robustness in \cite{Childs:01b}.

\section{How to model decoherence}\label{sec:decoherence}

No quantum system can be perfectly isolated from its surroundings and be viewed as perfectly closed. In the physical
world, degrees of freedom are usually interacting with many other degrees of freedom. In fact, the understanding of
this point is crucial for the explanation of why classical mechanics in the macroscopic world emerges out of the
microscopic operation of quantum mechanics. Even if we find quantum computer elements that interact only weakly with
the rest of the world (achievable most likely if they are themselves of atomic or near-atomic dimensions), for short
times the evolution will be unitary, but eventually even weak interactions will cause significant departure from
unitarity. Physical systems have a characteristic time for loss of unitarity, which is known in the field of mesoscopic
physics as the ``dephasing time''. It is often extremely short (for a table of dephasing times for various systems see
\cite{DiVincenzo:95a}), for example for the state of an electron traversing a gold wire at temperature less than 1K it
is of order $10^{-13}$ seconds.

We refer to the effects of noise due to unwanted coupling with the environment as decoherence\footnote{An unfortunate
confusion in terms has arisen with the word ``decoherence''. Historically it has been used to refer just to a phase
damping process - a specific type of noise - cf. e.g. Zurek \cite{Zurek:91a}. Zurek and others realized the unique role
played by phase damping in the transition from a quantum to a classical world. However, in the quantum computing
community by and large the term ``decoherence'' is now used to refer to {\em any noise process} in quantum
processing.}. An early treatise on quantum noise from a rather mathematical point of view is due to Davies
\cite{Davies:76a}. Caldeira and Leggett \cite{Caldeira:83a} in 1983 undertook one of the first and most complete
studies of an important model, the {\em spin-boson model}.

Within the context of quantum computers these studies were taken up by Unruh \cite{Unruh:95a} in 1995 and developed by
many others (e.g. Palma et al. \cite{Palma:96a}, Zanardi \cite{Zanardi:97b,Zanardi:98a}). Over the past few years work
on quantum computation has generated profound insights into the nature of decoherence.

\subsection{Hamiltonian Picture}\label{sec:Hamiltonian}
To model the dynamics of a register of qubits (quantum computer) with its surroundings we imagine the system immersed
into its environment (often called bath) and the whole (quantum register plus environment) as a closed system described
in a general way by the following Hamiltonian:
\begin{equation} \label{eq:generalH}
{\bf H}={\bf H}_{S}\otimes {\bf I}_{B}+{\bf I}_{S}\otimes {\bf H}_{B}+{\bf H} _{I},
\end{equation}
where ${\bf H}_{S}$ (${\bf H}_{B}$) [the system (bath) Hamiltonian] acts on the system (bath) Hilbert space ${\mathcal
H}_{S}$ (${\mathcal H}_{B}$), ${\bf I} _{S} $ (${\bf I}_{B}$) is the identity operator on the system (bath) Hilbert
space, and ${\bf H}_{I}$, which acts on both the system and bath Hilbert spaces ${\mathcal H}_{S}\otimes {\mathcal
H}_{B}$, is the interaction Hamiltonian containing all the nontrivial couplings between system and bath. In general $
{\bf H}_{I}$ can be written as a sum of operators which act separately on the system (${\bf S}_{\alpha }$'s) and on the
bath (${\bf B}_{\alpha }$'s):
\begin{equation}
{\bf H}_{I}=\sum_{\alpha }{\bf S}_{\alpha }\otimes {\bf B}_{\alpha }. \label{eq:HI-1}
\end{equation}
(Note that this decomposition is not necessarily unique.)

In the absence of an interaction Hamiltonian (${\bf H}_{I}=0$), the evolution of the system and the bath are separately
unitary: ${\bf U} (t)=\exp [-i{\bf H}t]=\exp [-i{\bf H}_{S}t]\otimes \exp [-i{\bf H}_{B}t]$ (we set $\hbar =1$
throughout). Information that has been encoded (mapped) into states of the system Hilbert space remains encoded in the
system Hilbert space if ${\bf H}_{I}=0$. However in the case when the interaction Hamiltonian contains nontrivial
couplings between the system and the bath, information that has been encoded over the system Hilbert space does not
remain encoded over solely the system Hilbert space but spreads out instead into the combined system and bath Hilbert
space as the time evolution proceeds.

Very often to describe decoherence in more specific contexts \cite{Unruh:95a,Palma:96a} it is convenient to model the
environment (the bath) as a mass-less scalar field, usually assumed to be in a thermal state (described by a density
matrix in Fock-space, the state space used to model {\em fields}. Its (infinite dimensional) Hilbert space is spanned
by
 \be \bigotimes_{\bf k} |i\ra_{\bf k}  \quad i \in \{0,1,2,\ldots \} \nonumber \ee
  where ${\bf k}$ labels the {\em modes}. On
each mode (factor in the tensor product) we have two operators, the {\em lowering operator} ${\bf b_k}$ given by ${\bf
b_k}|i\ra_{\bf k} =\sqrt{i}|i-1\ra_{\bf k}$ and the {\em raising operator} ${\bf b_k^\dagger}$ given by ${\bf
b_k^\dagger}|i\ra_{\bf k} =\sqrt{i+1}|i+1\ra_{\bf k}$. Note that ${\bf b_k^\dagger} {\bf b_k}|i\ra_{\bf k}=i|i\ra_{\bf
k}$, i.e. $|i\ra_{\bf k}$ is an {\em eigenstate} of the {\em number operator} ${\bf b_k^\dagger} {\bf b_k}$.

The quantum register is described by an arrangement of $n$ two-level systems (spins). This results in the {\em
spin-boson model}, where the bath Hamiltonian can be written as
\begin{eqnarray*}
{\bf H}_{B}=\sum_{k}\omega _{k}{\bf B}_{k}
\end{eqnarray*}
and, e.g., for the spin-boson Hamiltonian, ${\bf B}_{k}={\bf b} _{k}^\dagger {\bf b}_{k}$ \cite{Leggett:87a}, and ${\bf
b}_{k}^{\dagger }$, $ {\bf b}_{k} $ are respectively creation and annihilation operators of bath mode $k$. The
interaction Hamiltonian is given by
\begin{eqnarray} \label{eq:spinbosonH}
{\bf H}_{I}=\sum_{i=1}^{n}\sum_{\alpha =+,-,z}\sum_{k}g_{ik}^{\alpha }\sigma _{\alpha}^{i}\otimes {\tilde{{\bf
B}}}_{k}^{\alpha }+h.c.,
\end{eqnarray}
where $g_{ik}^{\alpha }$ is a coupling coefficient and $h.c.$ denotes the hermitian conjugate. In the spin-boson model
one would have ${\tilde{{\bf B}}}_{k}^{+}={\bf b}_{k}$, ${\tilde{{\bf B}}} _{k}^{-}={\bf b}_{k}^{\dagger }$ and
${\tilde{{\bf B}}}_{k}^{z}={\bf b} _{k}^{\dagger }+{\bf b}_{k}$. Thus $\sigma ^{i}_{\pm }\otimes {\tilde{{\bf B}
}}_{k}^{\pm }$ expresses a dissipative coupling (in which energy is exchanged between system and environment), and
$\sigma ^{i}_{z}\otimes {\ \tilde{{\bf B}}}_{k}^{z}$ corresponds to a phase damping process (in which the environment
randomizes the system phases, e.g., through elastic collisions).

\subsection{Operator Sum Picture}\label{sec:OSR}
The evolution of a quantum state in the entire space is unitary in a closed system of which we can observe and control
all parts. Very often, however, this is not the case: imagine for example that we perform a certain measurement and
then ``forget'' or lose the measurement outcome. As a result we know that the state has collapsed into some eigenstate
of of the measurement operator, but not into which one, and we will have to assign probabilities to each of them to
model the current state of the system. Take the example of a qubit $|\psi\ra=\alpha|0\ra + \beta |1\ra$ and the
measurement in the computational basis.
Performing this measurement and ``throwing away'' the result will leave the system in the state $|0\ra$ with
probability $|\alpha|^2$ and in the state $|1\ra$ with probability $|\beta|^2$. To describe this {\em mixture} of
possible states the {\em density matrix} formalism proved to be very useful: we write
\begin{equation}
\label{eq:rho1} \rho=|\alpha|^2 |0\ra \la 0| + |\beta|^2 |1\ra \la 1|=\left(\begin{array}{cc} |\alpha|^2 & 0 \\ 0 &
|\beta|^2 \end{array} \right)\nonumber
\end{equation}
Another way we can think about density matrices is to imagine that we have a (big) quantum system and can access only
part of it. To describe the quantum state of the accessible part (call it $A$), we have to average over the non-
accessible degrees of freedom of the system (part $B$). This is done by performing a complete measurement on system $B$
(mentally) and ``throwing away'' the outcomes (because we do not have access to them). Let us give the example of a
state of $2$ qubits $|\psi\ra_{AB}=\alpha|00\ra+\beta|11\ra$ where we only have access to the first qubit (part $A$).
If we (mentally) measure system $B$ in the computational basis, we obtain the density matrix from Eq. (\ref{eq:rho1}).

Let us proceed to the more general description of the statics and dynamics of open quantum systems, described by mixed
states:

\paragraph{States:} States in an $N$-dimensional Hilbert space ${\mathcal H}_N$ are given by density matrices $\rho$ such that:
\begin{itemize}
\item{$\rho$ is hermitian: $\rho^\dagger=\rho$} \item{$\rho$ is positive: $\forall |\psi\ra \in {\mathcal H}_N \quad
\la \psi|\rho|\psi\ra \geq 0$, which is equivalent to $\lambda_i \geq 0$, where $\lambda_i$ are the eigenvalues of
$\rho$ (this can be viewed as a statement about the positivity of probabilities of the pure states in the mixture).}
\item{$\rho$ has trace $1$ (this corresponds to the normalization of probabilities)}
\end{itemize}
Pure states $|\psi\ra$ of the system are associated with the density matrix $\rho_{pure}=|\psi\ra \la \psi|$. A general
mixed state is diagonalizable and can be written in its spectral decomposition as
\begin{equation}
\label{eq:rhodecomp} \rho=\sum_k p_k |\psi_k\ra \la \psi_k|\nonumber
\end{equation}
Note that there are in general many other ways to write $\rho$ in the above form if we allow for non-orthogonal states
in the decomposition. Each such decomposition $\{q_k, |\phi_k\ra\}$ is called an {\em ensemble realization} of $\rho$.
The ambiguity in the decomposition of $\rho$ manifests some loss of information, in the sense that the probabilistic
mixture $\rho$ could have arisen in a multitude of ways.

\paragraph{Dynamics:} To describe the evolution of an open system - and thus of decoherence -  we will immerse it into a closed system. The
evolution of a closed system is described by a unitary transformation, which translates to an effective dynamics of the
open system governed by completely positive (and trace-preserving) maps. Loosely speaking, these are maps that in
${\mathcal H}_A$ take density matrices to density matrices, with the additional property that if we extend the map to
bigger spaces ${\mathcal H}_{AB}$ by applying the map on $A$ and the identity map on $B$, then they still take density
matrices to density matrices. More precisely, an operator map $\Lambda$ is completely positive if $(I \otimes
\Lambda)(\rho) \geq 0$ whenever $\rho \geq 0$.

 The important point here is that according to {\em Kraus' Representation Theorem}
\cite{Kraus:83a} every completely positive trace preserving map can be written as
\begin{equation} \label{eq:krausrep}
\rho \rightarrow \sum_\mu M_\mu \rho M_\mu^\dagger \quad \mbox{with } \sum_\mu M_\mu^\dagger M_\mu=I
\end{equation}
where the $M_\mu$ are $N$-by-$N$ matrices ($N$ being the dimension of the Hilbert space).
 In particular this describes both the Hamiltonian and the Markovian dynamics, though in general it is often tedious to
 derive the form of the $M_\mu$ from the ${\bf S}_{\alpha}$ of the Hamiltonian picture
(Eq.~(\ref{eq:HI-1})). Note that contrary to the case of unitary evolution, general open system dynamics is not {\em
reversible}.

\subsection{Markovian Picture}\label{sec:markovian}

Another very powerful formalism to describe decoherence is the approach of {\em master equations}.  {\em Markovian
quantum dynamics} describes processes resulting from the interaction with a Markovian environment in the so called
Born-Markov approximation. The main objective is to describe the time-evolution of an open system with a differential
equation - the {\em Master equation} - which properly describes non-unitary behavior.

In fact it is not a priori obvious that there needs to be a differential equation that describes decoherence. Such a
description will be possible only if the evolution of the quantum system will be local in time ({\em Markovian}), i.e.
that the density operator $\rho(t+dt)$ is completely determined by $\rho(t)$. This is usually not the case because the
bath retains a memory of the state of $\rho$ at previous times for a while and can transfer it back to the system.

To obtain the Master equation in the Born-Markov approximation a common approach is to start with the Hamiltonian
description Eq.~(\ref{eq:generalH}) and use time-dependent perturbation theory (i.e. an expansion into time-series)
with careful truncation (cf. \cite{Carmichael:93a}).

A more axiomatic way, followed by Lindblad \cite{Lindblad:76a,Alicki:87a}, is to establish the most general linear
equation for density matrices. More precisely, by assuming that (i) the evolution of the system density matrix is a
one-parameter semigroup, (ii) the system density matrix retains the properties of a density matrix including ``complete
positivity'', and (iii) the system and bath density matrices are initially decoupled, Lindblad \cite{Lindblad:76a} has
shown that the most general evolution of the system density matrix ${\bf \rho }_{S}(t)$ (in a Hilbert space of
dimension $N$) is governed by the master equation
\begin{eqnarray}
{\frac{d{\bf \rho }}{dt}} ={\bf L}[\rho]&=&-i[{\bf H}_{S},{\bf \rho }]+{\frac{1}{2}}\sum_{\alpha ,\beta
=1}^{M}a_{\alpha \beta }\left( [{\bf F}_{\alpha },{\bf \rho }{\bf F} _{\beta }^{\dagger }]+[{\bf F}_{\alpha }{\bf \rho
},{\bf F}_{\beta }^{\dagger }]\right)\nonumber \\  &=& -i[{\bf H}_{S},{\bf \rho }]+{\frac{1}{2}}\sum_{\alpha ,\beta
=1}^{M}a_{\alpha \beta } {\bf L}_{\alpha,\beta}[\rho]. \label{eq:mastereq1}
\end{eqnarray}
Here ${\bf H}_{S}$ is the system Hamiltonian generating unitary evolution plus possible additional terms due to the
interaction with the bath - usually referred to as {\em Lamb-shift} -; the operators ${\bf F} _{\alpha }$ constitute a
basis for the $M$-dimensional space of all bounded operators acting on ${\mathcal H}_{S}$\footnote{they are often
called the {\em Lindblad operators} or the {\em quantum jump operators}}, and $a_{\alpha \beta }$ are the elements of a
positive semi-definite Hermitian matrix. We refer to the matrix $a_{\alpha\beta}$ as the {\em GKS matrix}.

Every such process described by Eq.~(\ref{eq:mastereq1}) corresponds to some interaction which, if applied for a
duration $t$, induces a quantum operation ${\mathcal E}_t$.  The class of quantum operations ${\mathcal E}_t$ forms a
Markovian semigroup, such that
\begin{equation}
  {\mathcal E}_s {\mathcal E}_t = {\mathcal E}_{s+t} \nonumber
\,.
 \end{equation}
  Here ${\mathcal E}_s {\mathcal E}_t$ denotes composition of the operations, i.e., ${\mathcal
E}_s \circ {\mathcal E}_t$.  Each Markovian semigroup in turn describes the dynamics resulting from some interaction
with a Markovian environment in the Born approximation.

Note that the Operator Sum Representation also describes Markovian dynamics, though it is in practice often difficult
to derive the $M_\mu$ (Eq. (\ref{eq:krausrep})) from the ${\bf F}_{\alpha}$ of the Markovian picture
(Eq.~(\ref{eq:mastereq1})).

To make our description of Markovian quantum dynamics concrete, we present some important examples of qubit noise
processes\footnote{For a review of these processes and their relevance to quantum information
  theory, see \cite{Nielsen:book}; \cite{Preskill:notes}.}.  We choose
the basis $\{F_\alpha\}$ to be the normalized Pauli operators $\frac{1}{\sqrt{2}} \{\sigma_x, \sigma_y, \sigma_z\}$,
and we write the density matrix of a qubit as \be \rho = \mtwo{\rho_{00}}{\rho_{01}}{\rho_{10}}{\rho_{11}}. \nonumber
\ee The first process, {\em phase damping}, acts on a qubit as \be
  {\mathcal E}^{\rm PD}_t(\rho)
     = \mtwo{\rho_{00}}{e^{-\gamma t} \rho_{01}}
            {e^{-\gamma t} \rho_{10}}{\rho_{11}}
\,, \nonumber \ee where $\gamma$ is a decay constant and $t$ is the duration of the process. The generator has a GKS
matrix with $a_{33}^{\rm PD} = {\gamma \over 2}$ and all other $a_{\alpha \beta}^{\rm PD} = 0$.  The second example is
the {\em depolarizing channel}, which acts on a qubit as \be
  {\mathcal E}^{\rm DEP}_t(\rho) =
    \mtwo{1+e^{-\tilde\gamma t}(\rho_{00}-\rho_{11}) \over 2}
         {e^{-\tilde\gamma t}\rho_{01}}
         {e^{-\tilde\gamma t}\rho_{10}}
         {1+e^{-\tilde\gamma t}(\rho_{11}-\rho_{00}) \over 2}
\,. \nonumber \ee Its GKS matrix has the nonzero elements $a_{11}^{\rm DEP}=a_{22}^{\rm DEP}=a_{33}^{\rm
DEP}=\tilde\gamma/4$. Our final example is {\em amplitude damping}, which acts on a qubit as \be
  {\mathcal E}^{\rm AD}_t(\rho) =
    \mtwo{\rho_{00}+(1-e^{-\Gamma t}) \rho_{11}}
         {e^{-\Gamma t/2} \rho_{01}}
         {e^{-\Gamma t/2} \rho_{10}}
         {e^{-\Gamma t} \rho_{11}}
\,. \nonumber \ee The GKS matrix $a_{\alpha \beta}^{\rm AD}$ is given by
 \be
     {\frac{\Gamma}{4}} \left( \begin{array}{ccc}{1}&{-i}&{0}\\
                             {i}&{1}&{0}\\
                             {0}&{0}&{0} \end{array} \right)
\,. \ee

\section{The Error-model}\label{sec:error-model}

The underlying key assumption for efficient usage of quantum error-correcting codes is the {\em independent error
model}. Intuitively, if a noise process acts independently on the different qubits in the code, then provided the noise
is sufficiently weak, error-correction should improve the storage fidelity of the encoded over the unencoded state.

Mathematically, the assumption of independent errors can be retraced in each of the decoherence pictures introduced in
Sec. \ref{sec:decoherence}. In the Hamiltonian picture we can rewrite Eq.~(\ref{eq:spinbosonH}) as
\begin{eqnarray*}
{\bf H}_{I}=\sum_{i=1}^{K}\sum_{\alpha =x,y,z}\sum_{k}\sigma _{i}^{\alpha }\otimes {\bf B}_{ik}^{\alpha },
\end{eqnarray*}
where ${\bf B}_{ik}^{z}\equiv {\tilde{{\bf B}}}_{k}^{z}$ and ${\bf B} _{ik}^{x}$ ,${\bf B}_{ik}^{y}$ are appropriate
linear combinations of ${\ \tilde{{\bf B}}}_{k}^{+}$ and ${\tilde{{\bf B}}}_{k}^{-}$:
\begin{eqnarray*}
{\bf B}_{ik}^{x} &=&\frac{1}{2}\left( g_{ik}^{-}{\tilde{{\bf B}}}
_{k}^{-}+g_{ik}^{+}{\tilde{{\bf B}}}_{k}^{+}\right) \\
{\bf B}_{ik}^{y} &=&\frac{i}{2}\left( g_{ik}^{-}{\tilde{{\bf B}}} _{k}^{-}-g_{ik}^{+}{\tilde{{\bf B}}}_{k}^{+}\right)
\end{eqnarray*}
i.e. all system components can be expressed in terms of tensor products of the single qubit {\em Pauli matrices}. If we
expand the evolution to first order in time and assume that the error-rates $g_{ik}$ are independent we will get an
operator sum representation (OSR) (cf. Eq.~(\ref{eq:krausrep})) where each term is a linear combination of the Pauli
matrices (see \cite{Lidar:99b} for a recent derivation). In the Markovian formulation of noise (cf.
Eq.~(\ref{eq:mastereq1})) the independent error model assumes that each of the ${\bf F}_\alpha$ affects only one of the
qubits and that the ${\bf F}_\alpha$ are not correlated. Higher order correlations are taken into account by using a
code that is suitably constructed for the particular error-model. Therefore the theory of QECCs has focused on
searching for codes that make quantum information robust against 1, 2,... or more erroneous qubits, as this is the most
reasonable model when one assumes spatially separated qubits with their own local environments.  {\em Detection and
correction procedures must then be implemented at a rate higher than the intrinsic error rate}.

From the linear decomposition of the error operators in the OSR or the master equation it follows that QECC-schemes
need to be able to correct only a discrete set of errors, namely those generated by the Pauli-group\footnote{In the
context of error-correction the Pauli matrices $\sigma_{x,y,z}$ are often denoted by ${\bf X,Y,Z}$ respectively.}.
Intuitively we can imagine that the error process acting on one qubit puts the quantum state into a superposition of
one of the four possible discrete errors ($I_2,\sigma_{x,y,z}$) and the error-detection and correction procedure
collapses the state into one of these errors and then corrects as needed. This intuition can be made formal
\cite{Nielsen:book,Knill:97b}: it is possible to decompose the the operators $M_\mu$ that appear in Eq.
(\ref{eq:krausrep}) into a basis of tensor products of the Pauli matrices. It can then be seen, when deriving the OSR
representation from the Hamiltonian picture, that to first order in the noise rate the noise process gives terms with a
single non-identity matrix (single qubit error). The core message is again that quantum codes need to account only for
a discrete {\em linear basis} of all possible errors.

\section{Stabilizer Codes}\label{sec:stabilizer}

Stabilizer codes, also known as {\em additive} codes, are an important subclass of quantum codes. The stabilizer
formalism provides an insightful tool to quantum codes and fault-tolerant operations. It was developed by Gottesman in
1996 \cite{Gottesman:97a,Gottesman:97b}. We will not outline the full formalism here but rather describe only the key
elements. A full treatment can be found in \cite{Gottesman:PHD}.

The powerful idea behind the stabilizer formalism is to look at the set of group elements that {\em stabilize} a
certain code and to work with this stabilizer instead of directly with the code. In the framework of QECCs, the
stabilizer permits on the one hand to identify the errors the code can detect and correct. It also links quantum codes
to the theory of classical error correcting codes in a transparent fashion. On the other hand it also allows one to
find a set of universal, fault-tolerant gates.

An operator ${\bf S}$ is said to stabilize a code ${\mathcal C}$ if
\begin{equation}
|\Psi \rangle \in {\mathcal C}\quad {\rm iff}\quad {\bf S}|\Psi \rangle =|\Psi \rangle \quad \forall {\bf S}\in
{\mathcal S}.  \label{eq:stabicond}
\end{equation}
The set of operators $\{{\bf S}\}$ form a group ${\mathcal S}$, known as the stabilizer of the code
\cite{Gottesman:97b}. Clearly, ${\mathcal S}$ is closed under multiplication. In the theory of QECC the underlying
group is the Pauli-group, the stabilizers are subgroups of the Pauli-group (tensor products of ${\bf I,X,Y,Z}$ ). Since
any two elements of the Pauli group either commute or anti-commute, the stabilizer, in this case is always {\em
Abelian}. The code is thus the common eigenspace of the stabilizer elements with eigenvalue $1$. {\em Additive} codes
are completely characterized by their stabilizer $\mathcal S$. The stabilizer $\mathcal S$ can be given by a set of
generators which span the stabilizer group via multiplication.

We define the {\em centralizer} of $\mathcal S$ to be the set of elements $e$ in the Pauli group that commute with
every element in $\mathcal S$, i.e. $e{\bf S}={\bf S}e$ for all ${\bf S} \in {\mathcal S}$. In case of the Pauli group
it coincides with the {\em normalizer} of $\mathcal S$ - the set of elements $E$ in the Pauli-group with $E {\bf S}
E^\dagger \in {\mathcal S}$ for all ${\bf S} \in {\mathcal S}$. We will denote it by $N({\mathcal S})$ and call it
normalizer throughout. Note that the normalizer contains the stabilizer $\mathcal S$ itself.

Recall that in the theory of QECCs the error process $\mathcal E$ can be expanded in terms of the error basis ${\bf E}$
which is a subgroup of the Pauli group. In particular ${\bf E}_\alpha \in {\bf E}$ either commutes or anti-commutes
with elements in the stabilizer. This allows us to recast the QECC-condition Eq.~(\ref{eq:QECCcond1}) in the stabilizer
formalism as follows

\noindent {\em QECC-conditions: A quantum code ${\mathcal C}$ with stabilizer $\mathcal S$ is an $\mathcal
E$-correcting QECC if for all ${\bf E}_\alpha,{\bf E}_\beta \in {\bf E}$ one of the following holds:

(1) There is an ${\bf S} \in {\mathcal S}$ that anti-commutes with ${\bf E}_\alpha^\dagger {\bf E}_\beta$

(2) ${\bf E}_\alpha^\dagger {\bf E}_\beta \in {\mathcal S}$.}

This clearly implies the QECC conditions Eq.~(\ref{eq:QECCcond1}), since in the case of (1) $\la \Psi_i|{\bf
E}_\alpha^\dagger {\bf E}_\beta|\Psi_j\ra=\la \Psi_i|{\bf E}_\alpha^\dagger {\bf E}_\beta {\bf S}|\Psi_j\ra=-\la
\Psi_i|{\bf S}{\bf E}_\alpha^\dagger {\bf E}_\beta|\Psi_j\ra=0$ and in the case of (2) $\la \Psi_i|{\bf
E}_\alpha^\dagger {\bf E}_\beta|\Psi_j\ra=\la \Psi_i|\Psi_j\ra=\delta_{ij}$. In particular this implies that the matrix
element $c_{\alpha \beta}$ is either $0$ in the case of (1) or $1$ in the case of (2). Conditions (1) and (2) can be
reformulated succinctly as: ${\bf E}_\alpha^\dagger {\bf E}_\beta \notin N({\mathcal S})-{\mathcal S}$.

The nine bit Shor code in Eq. (\ref{eq:Shorcode}) is a stabilizer code. The set of its stabilizers is generated by
 \begin{align*}
 Z_1Z_2 \quad \quad Z_2Z_3 \quad \quad Z_4Z_5  \quad&\quad Z_5Z_6 \quad \quad Z_7Z_8\quad \quad Z_8Z_9\nonumber \\
 X_1X_2X_3X_4X_5X_6 \quad X_1X_2X_3&X_7X_8X_9 \quad X_4X_5X_6X_7X_8X_9
 \end{align*}
Note that these generators correspond to the measurements to detect bit flip (the $Z$ generators) and phase flip (the
$X$ generators) errors. Indeed, for instance a bit flip error on the first qubit anticommutes with $Z_1Z_2$.

The generators for the stabilizer of the Steane code in Eq. (\ref{eq:Steane}) on the $7$ qubits are the following:
\begin{align*}
 IIIZZZZ \quad \quad  \quad \quad  IIIXXXX \nonumber \\
 IZZIIZZ \quad \quad  \quad \quad  IXXIIXX \nonumber \\
 ZIZIZIZ \quad \quad  \quad \quad  XIXIXIX.
 \end{align*}
Note how the positions of the $Z$ and $X$ correspond to the parity check matrix $H$ in Eq. (\ref{eq:Hamming}). The fact
that the code is self dual is seen in the symmetry between the $Z$ and the $X$. Note that the change of basis from
$\ket{0}$, $\ket{1}$ to the $\ket{\pm}$ basis, which is implemented by a conjugation with the Hadamard transform $H$ on
each bit, transforms all $Z$ into $X$ and vice versa ($HXH=Z$ and $HZH=X$).

The smallest possible quantum code to protect against single qubit errors, the $5$-qubit code \cite{Laflamme:96a}, has
the following (shift invariant) stabilizer
\begin{align*}
  XZZXI \nonumber \\
  IXZZX \nonumber \\
  XIXZZ \nonumber \\
  ZXIXZ .
 \end{align*}


The stabiliser formalism allows to derive fault-tolerant computation in a convenient way. A key ingredient are encoded
gates that transform encoded states, without decoding them, which would expose them to noise without protection. For
universal quantum computation it is sufficient to show how to implement a universal {\em discrete} set of gates
fault-tolerantly.

The key insight to fault-tolerant gates is that these encoded gates should only take encoded states to valid encoded
states, without leaving the code-space. For an encoded gate $G$ this means that after application of $G$ to a state
stabilized by all elements of $\mathcal S$ the resulting state must still be stabilized by $\mathcal S$ (see
Eq.~(\ref{eq:stabicond})) \be G|\Psi\ra \in {\mathcal C} \quad \leftrightarrow \quad SG|\Psi\ra=GS|\Psi\ra \ee or in
other words over the code-space $G$ commutes with all elements of $\mathcal S$. In the case of QECCs and the Pauli
group this means that $G$ is in the {\em normalizer} $N({\mathcal S})$ of ${\mathcal S}$.

The normalizer allows one to easily identify encoded logical operations on the code. It can be shown that for large
classes of stabilizer codes a universal gate-set can be implemented either because the encoded gates are {\em
transversal}, i.e. they affect only one qubit per block, or in connection with state-preparation of special states and
measurement.

To fault-tolerantly {\em measure} on encoded states  ancilla-state are employed in a procedure like the
following\footnote{This procedure may differ from case to case, here we only give an example for illustration.}:
Suppose we wish to measure the encoded qubit in the encoded computational basis. The ancilla is prepared in the state
$|0_L\ra$. Then we perform an {\em encoded} $CNOT$ from the encoded qubit to be measured to the ancilla. We then
measure the ancilla state in the computational basis, which gives us a non-destructive measurement of the encoded qubit
in the encoded computational basis which is tolerant of possible errors in the encoded qubit. To prevent possible
uncontrolled error-propagation caused by an incorrectly prepared ancilla, we prepare multiple $|0_{L}\rangle $-ancillas
and apply $CNOT$'s between the DFS state to be measured and each ancilla. Together with majority voting this provides a
fault-tolerant method for measuring $\bar{Z}$ \cite{Gottesman:97a}.

The stabilizer formalism also provides an easy framework for fault-tolerant encoded state preparation and decoding, as
it turns out that only transversal measurements in the Pauli-Z-basis are needed for both. For a detailed account of
fault-tolerant computation with stabilizer codes see Gottesman \cite{Gottesman:PHD}.

\section{Noise model for Decoherence-Free Subsystems}\label{sec:DFS}

To derive the model of {\em collective} noise that applies to decoherence-free subsystems, we will work with the
Hamiltonian picture (see App. \ref{sec:Hamiltonian}), following \cite{Zanardi:97a}. We use the interaction Hamiltonian
Eq. (\ref{eq:spinbosonH}). {\em Collective decoherence} is the case where the coupling constants do not depend on the
qubit, i.e. $g^\alpha_{i{\bf k}}=g^\alpha_{{\bf k}}$. This allows to rewrite the Hamiltonian in terms of the operators
 \be \label{eq:Salpha} S_\alpha=\sum_{i=1}^n \sigma_\alpha^i \ee as
 \be H=\omega_0 S_z + \sum_{\bf k} \omega_{\bf k}b_{\bf k}^\dagger b_{\bf k}+\sum_{\alpha=\pm,z}  S_\alpha \otimes \sum_{\bf
k} g^\alpha_{\bf k} {\tilde {\bf B}}_{\bf k}^\alpha + h.c. \nonumber
 \ee
 The crucial observation is now that if we start the
system in a common eigenstate with the same eigenvalue of all the $S_\alpha$ with the bath in an eigenstate of $H_B$
then the evolution will be completely {\em decoupled}. The condition for decoherence-free subspaces (DFS) is
 \be \label{eq:DFSham}
S_\alpha |\Psi\ra = c_\alpha |\Psi\ra \quad \forall |\Psi\ra \in DFS
 \ee
Dynamical symmetry allows for unitary evolution of a subspace while the remaining part of the Hilbert space gets
strongly entangled with the environment. This is true for arbitrary coupling strength. The form of noise where all
three $S_\alpha$, $\alpha \in \{x,y,z\}$, come into play is now called {\em strong collective decoherence}, if there is
only coupling to one of the $S_\alpha$ the noise is called {\em weak collective decoherence}.

It is also possible to study collective noise in the Markovian picture (see App. \ref{sec:markovian}). We use Eq.
(\ref{eq:mastereq1}), where ${\bf L}_D$ gives the non-unitary ``coupling term\footnote{Note that the coupling of a
system with an environment might also change the unitary part of the evolution of $-i[{\bf H},\rho]$ by introducing an
additional term to the system Hamiltonian, called the {\em Lamb-shift}.}. The decoherence-free condition ${\bf
L}_D[\rho]=0$ implies that
 \be  {\bf F}_\alpha |\Psi\ra = c_\alpha |\Psi\ra \quad \forall |\Psi\ra
\in DFS \nonumber
 \ee As before the decoherence-free states are common eigenstates of the operators ${\bf F}_\alpha$. In case of
 collective decoherence (symmetry of all the qubits), the ${\bf F}_\alpha$ are exactly the $S_\alpha$ of Eq.
 (\ref{eq:Salpha}), and it is possible to show that the unitary term of the Master-equation does not affect the
 DFS to first order.

This line of reasoning can be generalized to decoherence-free {\em subsystems}. The $S_\alpha$ act on the system space.
We can study the irreducible representations of this action and identify irreducible subspaces. For each irreducible
representation there will be one or several irreducible subspaces on which the $S_\alpha$ act in the same way. The one
dimensional subspaces will only get a phase factor, they correspond to the decoherence-free subspaces of Eq.
(\ref{eq:DFSham}). But the other subspaces are not lost for our purposes. Any irreducible subspace can be used to
encode information, because the action of the noise operators $S_\alpha$ will keep the state within the subspace. Even
though the state will change, its subspace will identify the encoded information. The number of irreducible subspaces
corresponding to the same irreducible representation gives the number of different code words we can use.

The irreducible representations corresponding to the operators associated with strong collective decoherence (Eq.
(\ref{eq:Salpha})) have been studied widely in physics, as they correspond to the {\em angular momentum} operators.

In the case of two qubits there is a single common eigenstate of the $S_\alpha$,  $\alpha \in \{x,y,z\}$, the singlet
state \be |\Phi\ra=\frac{1}{\sqrt{2}}(|0\ra|1\ra-|1\ra|0\ra). \nonumber\ee

For three qubits there is no one-dimensional irreducible representation of the $S_\alpha$, but there are two
(identical) two-dimensional irreducibles subspaces into which we can encode as

 \be
|0\rangle_{code} =\left\{
\begin{array}{l}
|{\frac{1}{\sqrt{2}}}\left( |010\rangle -|100\rangle \right) \\
|{\frac{1}{\sqrt{2}}}\left( |011\rangle -|101\rangle \right)
\end{array}
\right. \quad \quad |1\rangle_{code} =\left\{
\begin{array}{l}
{\frac{1}{\sqrt{6}}}\left( -2|001\rangle
+|010\rangle +|100\rangle \right) \\
{\frac{1}{\sqrt{6}}}\left( 2|110\rangle -|101\rangle -|011\rangle \right)
\end{array}\nonumber
\right.
 \ee
Increasing the number of qubits the number of identical irreducible subspaces grows favorably, so that it is possible
to encode at a good rate. For more references on this and the theory of fault-tolerant computation on such systems, see
Sec. \ref{sec:reading}.

\bibliographystyle{alpha}

\newcommand{\etalchar}[1]{$^{#1}$}

\end{document}